\documentclass[useAMS,usenatbib]{mn2e}
\def\be{\begin{equation}}
\def\ee{\end{equation}}
\newcommand{\Msun}{M$_{\sun}$}
\newcommand{\Msunsp}{M$_{\sun}$ }
\usepackage{graphicx}
\usepackage{epsfig}
\usepackage{amsmath}
\usepackage{subfig}
\title[Angular Momentum Evolution in Dark Matter Halos]{Angular Momentum Evolution in Dark Matter Halos}
\author[Book et al.]{Laura G. Book\thanks{E-mail:
lbook@caltech.edu}$^{1}$, Alyson Brooks$^{1}$, Annika H.\ G.\ Peter$^{1}$, Andrew J.\ Benson$^{1}$,\newauthor Fabio Governato$^{2}$\\
$^{1}$Mail Code 350-17, California Institute of Technology, Pasadena, CA 91125\\
$^{2}$Astronomy Department, University of Washington, Box 351580, Seattle, WA 98195}

\begin{document}

\pagerange{\pageref{firstpage}--\pageref{lastpage}} \pubyear{2010}

\maketitle

\label{firstpage}

%
%
\begin{abstract}

 We have analyzed high resolution N-body simulations of dark matter halos, focusing specifically on the evolution of angular momentum. We find that not only is individual particle angular momentum not conserved, but the  angular momentum of radial shells also varies over the age of the Universe by up to factors of a few. We find that torques from external structure are the most likely cause for this distribution shift. Since the model of adiabatic contraction that is often applied to model the effects of galaxy evolution on the dark-matter density profile in a halo assumes angular momentum conservation, this variation implies that there is a fundamental limit on the possible accuracy of the adiabatic contraction model in modeling the response of DM halos to the growth of galaxies.

\end{abstract}
\begin{keywords}
galaxies:haloes, galaxies:kinematics and dynamics
\end{keywords}

%
%
%
%
%
%
\section{Introduction}  \label{sec:intro}

Recent years have seen much progress in our understanding of the growth of dark-matter halos from initial density fluctuations and the characterization of their properties. For example, it has been possible to find analytic results to describe the merger trees of dark matter-only systems \citep{Press74,Sheth02,benson_self-consistent_2005,zhang_to_2008,neistein_constructing_2008,parkinson_generating_2008}. Additionally, high-resolution N-body simulations containing only cold (i.e. non-relativistic at the epoch of kinetic decoupling) dark matter have been performed \citep{Springel05, Kuhlen08, boylan-kolchin2009, Stadel09, Klypin10}, leading to detailed knowledge of the structure of cold dark-matter halos, such as the fact that halos have an approximately universal radial density profile \citep{NFW97,navarro2004,navarro2010} and are generally triaxial in shape \citep{Dubinski91}.

There is currently a great deal of interest in the angular momentum of particles in dark matter halos, since it is this angular momentum, when transferred to baryons, that creates the disks of galaxies. It is also interesting to look at this topic in the light of the adiabatic contraction (AC) model, which is used to model how the condensation of baryons during the formation of a galaxy affects the mass profile of its host halo \citep[e.g.,][]{Mo98,Cole00,Somerville08,Benson10}. In this model, it is assumed that the gravitational potential of the system changes very slowly, so that it can be approximated as adiabatic. Further, two simplifying assumptions are usually employed to calculate the density increase resulting from the growth of galaxies at centres of dark-matter halos using AC.  First, the gravitational potential of the system (including contributions from both the dark-matter halo and the galaxy) is assumed to be spherically symmetric and the orbits of particles are presumed to be circular, such that the angular momentum of particles is conserved.

The back reaction of the evolution of those galaxies on the dark matter is poorly understood.  The properties of the galaxy population and of the dark-matter response appear to depend quite strongly on the star-formation prescription and strength of feedback processes, among other things, but a physically-motivated and vetted mapping between baryonic physics and the evolution in the dark-matter phase-space density is lacking \citep{Cole00,Gnedin04,Gustafsson06,romano-diaz2008b,romano-diaz2009,Somerville08,Abadi09,Tissera09,Benson10,Duffy10,Governato10}.

The development of an accurate model of this back reaction is required to accurately compare theoretical predictions of galaxy formation, both in the context of canonical cold-dark-matter and alternative dark-matter cosmologies with observations.  A number of observed galaxy properties, such as the rotation curves of disk galaxies and the associated Tully-Fisher relation \citep{tully1977}, depend on the gravitational potential of both baryons and dark matter.  In particular, the Tully-Fisher relation is frequently used as a constraint on galaxy evolution processes in models of galaxy formation \citep{Cole00, Hatton03, DeLucia04, Croton06, Monaco07, dutton_revised_2007, Somerville08, Dutton09, Benson10}.  Moreover, different dark-matter candidates are expected have noticeably different distributions in dark-matter halos in the absence of baryons \citep[e.g.,][]{spergel2000,abazajian2001,feng2003,kaplinghat2005}; it is not clear how those distributions will change as a result of baryonic physics.  Observed density profiles sometimes appear consistent with cold-dark-matter predictions in the absence of baryons, and sometimes do not \citep{simon2005,kuzio2008,deblok2008,padmanabhan2004,newman2009,treu2004,gavazzi2007,Schulz09}.  

In the absence of a physically-vetted predictive model for the impact on baryonic physics processes on dark-matter distributions in halos, variations on the adiabatic contraction (AC) model are often applied to compare galaxy evolution and dark-matter theories with observations \citep[e.g.,][]{Mo98,Cole00,Somerville08,Benson10}.  In its most general form, AC applies if the halo potential is only slowly evolving and is integrable, such that the Hamiltonian and the distribution function (DF) of particles are fully described by action variables.  If the time scale for changes in the potential is long compared to the typical dynamical time of particles in the halo, the action variables are invariant under changes in the potential and the DF (when expressed in those action variables) is therefore also invariant with time \citep{Binney08}.  However, transforming from a DF to a spatial distribution is non-trivial, and so simple approximations to the true AC model are often used instead.  The two most common simplifying assumptions are that the gravitational potential (from both the dark matter and the galaxy) is spherically symmetric, such that the magnitude of the angular momentum and the radial action are the relevant action variables; and that either all orbits are circular such that angular momentum is the only non-zero and non-infinite-period angle variable \citep{Blumenthal86}.  Occasionally, the assumption of circular orbits is swapped in favor of choosing a variant of the radial action as a conserved quantity \citep{Blumenthal86,Gnedin04}.  One can thus analytically calculate the final mass profile of the dark matter given the growth of the galaxy.  Although these simple models have a mixed track record of matching observations and hydrodynamic simulations of galaxy evolution, they are currently the only predictive models for the effects of galaxy evolution on the dark-matter profile in halos \citep{treu2004, gavazzi2007, Schulz09, simon2005, deblok2008, kuzio2008, Dutton09, Gustafsson06, romano-diaz2008b, romano-diaz2009, Abadi09, Governato10,Tissera09,Duffy10}.

In this work, we examine two key assumptions of the AC model in the context of dark-matter-only simulations of galaxy-mass halos: the adiabaticity of the evolution of the gravitational potential, and the angular momentum distribution of dark-matter particles halos. Specifically, we will investigate the extent to which angular momentum is invariant for individual particles and subsets of particles, as is assumed in the spherically-symmetric model that is generally applied (and which will be true even if particle orbits are non-circular). Invariance may be broken for a number of reasons, including a break down of the adiabatic assumption, the non-sphericity of the halo potential and torques from the external mass distribution.  Even if the angular momentum of individual particles is not conserved due to the triaxiality of the halo density profile, AC might be applicable if the \emph{distribution} of angular momenta of all particles were invariant with time.  If the angular momentum distribution varies with time then the simple model of AC cannot work precisely for even the simplest galaxy-evolution models.  

Note that, as we are working with dark-matter-only simulations, we set only lower limits on the level to which AC is not applicable in halos with both dark matter and baryons, as the inclusion of baryonic physics is likely to exacerbate these effects.  The observed angular momentum distribution of baryons in galaxies has been shown to deviate significantly from that expected based on simulations  \citep{vandenBosch01,vandenBosch02,vandenBosch03}, so we do not expect the angular momentum distribution that we measure to be representative of the baryonic distribution of angular momenta.  However, the extent of non-conservation of angular momentum in dark-matter-only simulations is likely to be less than that when baryons are added, since baryons and dark matter can exchange angular momentum.  Hence, in this paper, we examine the accuracy of the adiabatic assumption and the level to which the angular momentum distribution changes with time in dark-matter-only simulations, as this places a limit on the possible accuracy of the AC model.  Our intention is not to provide a precise quantification of this limit, but merely to highlight its existence and provide an approximate measure of its magnitude.

This paper is organized as follows: in \S\ref{sec:sim}, we describe the simulations and halos that we analyze, and we describe the particle subsets that we use in the paper in \S\ref{sec:subsets}. We show how the average angular momentum of halo particles with respect to the centre of mass is evolving in \S\ref{sec:evol}, and analyze the extent to which the adiabatic approximation is valid in the halos in \S\ref{sec:adiabatic}. In \S\ref{sec:Levol} we present the evolution of the average angular momentum of halo particles with respect to the centre of the halo, and we present the evolution of the angular momentum distribution in \S\ref{sec:Distrib}. Finally, we examine the causes of this evolution in \S\ref{sec:why}, and discuss our results in \S\ref{sec:disc}.

\section{Simulations}  \label{sec:sim}

The dark-matter halos used in this study were simulated with the code PKDGrav \citep{Stadel01} using a \emph{Wilkinson Microwave Anisotropy Probe} (WMAP) three-year cosmology \citep[$\Omega_{m}$ = 0.24, $\Omega_{\Lambda}$ = 0.76, H$_0$ = 73 km/s, $\sigma_{8}$ = 0.77;][]{spergel2007}. These four halos were originally chosen from a low-resolution volume of 50 Mpc on a side, and selected to span a range of merger histories and spin values at roughly the mass of the Milky Way halo. Each halo was then re-simulated using the volume renormalization technique \citep{Katz93}. This approach creates successively finer resolution layers around the halo of interest, allowing for high resolution on one halo while maintaining the large-scale structure (from the original 50 Mpc box) at lower resolution. Importantly, the large-scale structure can deliver tidal torques and angular momentum to the halo.  

Table \ref{simsum} lists properties of each of the simulated halos.  Three of the four halos have been presented at similar resolution, but including gas through smoothed particle hydrodynamics (SPH), in previous papers \citep[e.g.,][]{Brooks09, Pontzen09, Read09, Zolotov09, Governato09}. As examined in \citet{Read09} and \citet{Governato09}, halo h258 has an approximately binary merger that occurs at z $\sim$1, but exhibits a very quiescent evolution afterward. \citet{Zolotov09} showed that h277 has a fairly quiescent merger history back to z $\sim$3, while h285 experiences a large number of minor mergers all the way to redshift 0, despite not having a major merger since high redshift.  Halo h239 is presented here for the first time. It has a continually active merger history, both major and minor, until z~$\sim$~0.5.  

The mass resolution of the particles that make up these halos is 1.2$\times$10$^{6}$ \Msun, with a spline force softening of 350~pc. At each output time step ($\sim$80 Myr), high resolution halos with more than 64 particles \citep[above which the mass function converges,][]{Reed03, Governato07} in the volume are identified using AHF \citep[AMIGA's Halo Finder,][]{Gill04, Knollmann09}. AHF adopts results from \citet{Gross97}, calculating the overdensity assuming a spherical top hot
collapse, under the assumption that the halo has just virialized.  The
definition for $\delta_{vir}$ differs from its value as defined in \citet{Eke96} by the factor $\Omega(z)*(1+\delta_{vir})$. Thus $\Delta_c$, the value for which $\rho_{vir} = \Delta_c \rho_{crit}$, is $\sim 100$ at $z\,=\,0$. We follow the main halo through time by identifying the most massive progenitor at high redshift.

\begin{table}
\centering
\begin{minipage}{140mm}
\caption{ Simulated Galaxy Properties\label{simsum} }
\begin{tabular}{@{}lcccc@{}}
\hline \hline
simulation & $M_{\rm vir}$ [\Msun] & $\lambda$\footnote{Global spin parameter as defined in \citet{Bullock01}.} & z$_{\rm LMM}$ \footnote{Redshift of last major merger.} & N within R$_{\rm vir}$ \footnote{Number of dark matter particles within the virial radius at $z=0$.}\\
\hline
H239 & 9.3$\times$10$^{11}$ & 0.01 & 1.25 & 7.6$\times$10$^{5}$ \\
H258 & 8.2$\times$10$^{11}$ & 0.03 & 1.25 & 6.7$\times$10$^{5}$ \\
H277 & 7.2$\times$10$^{11}$ & 0.03 & 2.5  & 5.9$\times$10$^{5}$ \\
H285 & 7.4$\times$10$^{11}$ & 0.02 & 3.75 & 6.1$\times$10$^{5}$ \\
\hline
\end{tabular}
\end{minipage}
\end{table}

\subsection{Particle Subsets}  \label{sec:subsets}

There are several different subsets of particles whose properties we analyze here. We take two different approaches to selecting particles; in the first case, we choose particles based on their radius relative to the halo centre at $z=0$  and follow these same particles back through the simulation (Lagrangian selection), while in the second case we choose particles based on their radius at each time step (Eulerian selection). We use the Lagrangian method to highlight the evolution over the course of the simulation of those particles that will be at a certain radius at $z=0$. The Eulerian approach is complementary, as it shows how the particles at a certain radius at each timestep are related.  This is relevant in the context of galaxy evolution, as the baryons condense to a specific region of physical space.

We also use two different methods of averaging the angular momenta of particles, one adding them as vectors, and a second simply adding their magnitudes. These methods also highlight different features of the angular momentum distribution. The vector addition of angular momenta allows us to see the extent to which a given set of particles have their angular momenta aligned, as a decrease in angular momentum may represent a mixing of angular momentum directions as well as a change in magnitude. Thus, the vector-averaged angular momentum of a radial shell of particles is essentially telling us about the evolution of the `spin' of that shell. In contrast, adding the magnitudes allows us to single out only the change in the magnitude of particle angular momentum, and therefore tells us about changes in the particle orbits. It is this quantity which is relevant for the distribution function and adiabatic invariance.

\section{Halo Evolution}  \label{sec:evol}

\begin{figure*}
\centering
\subfloat[][]{\includegraphics[width=0.4\textwidth]{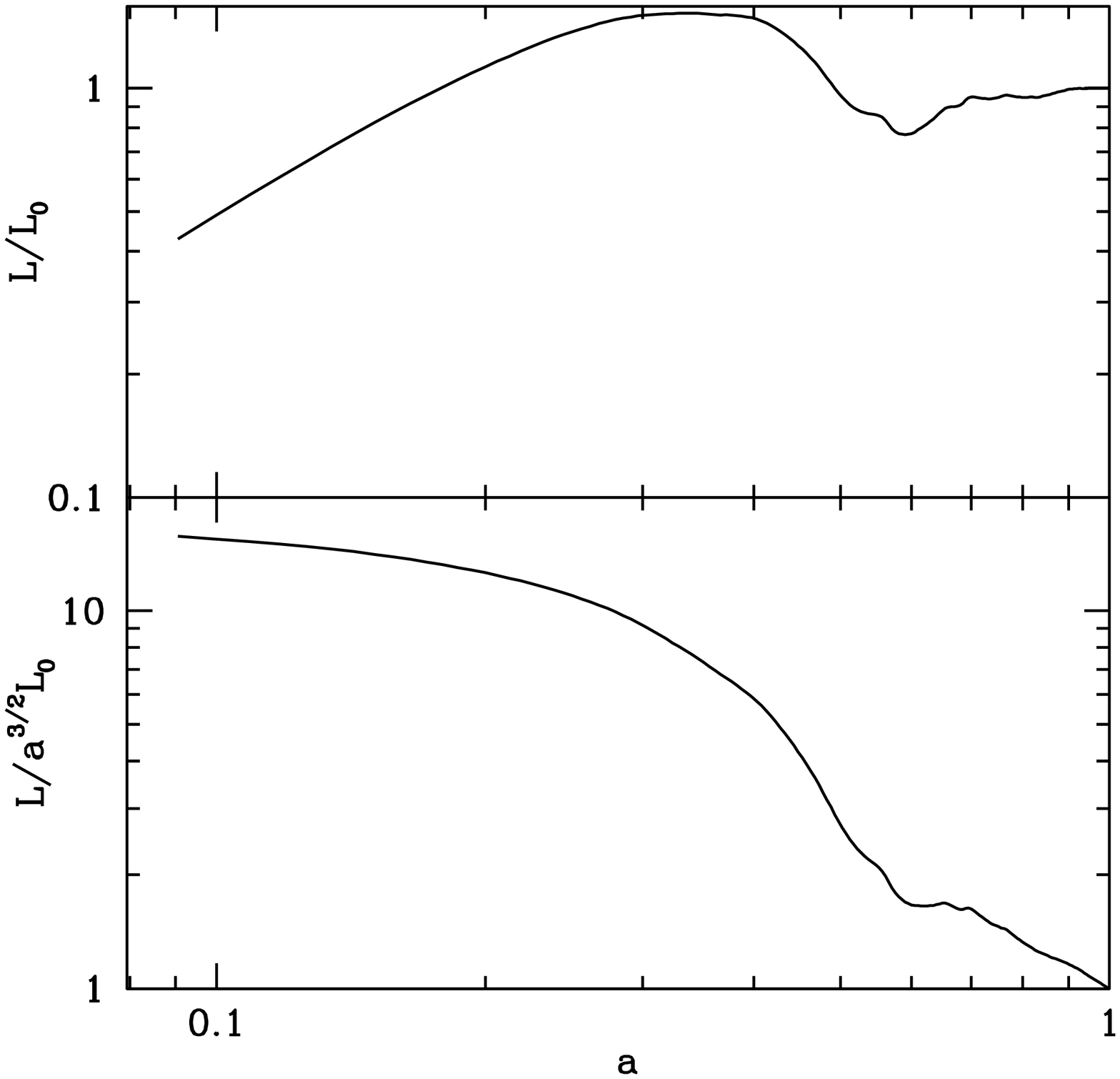}}
\subfloat[][]{\includegraphics[width=0.4\textwidth]{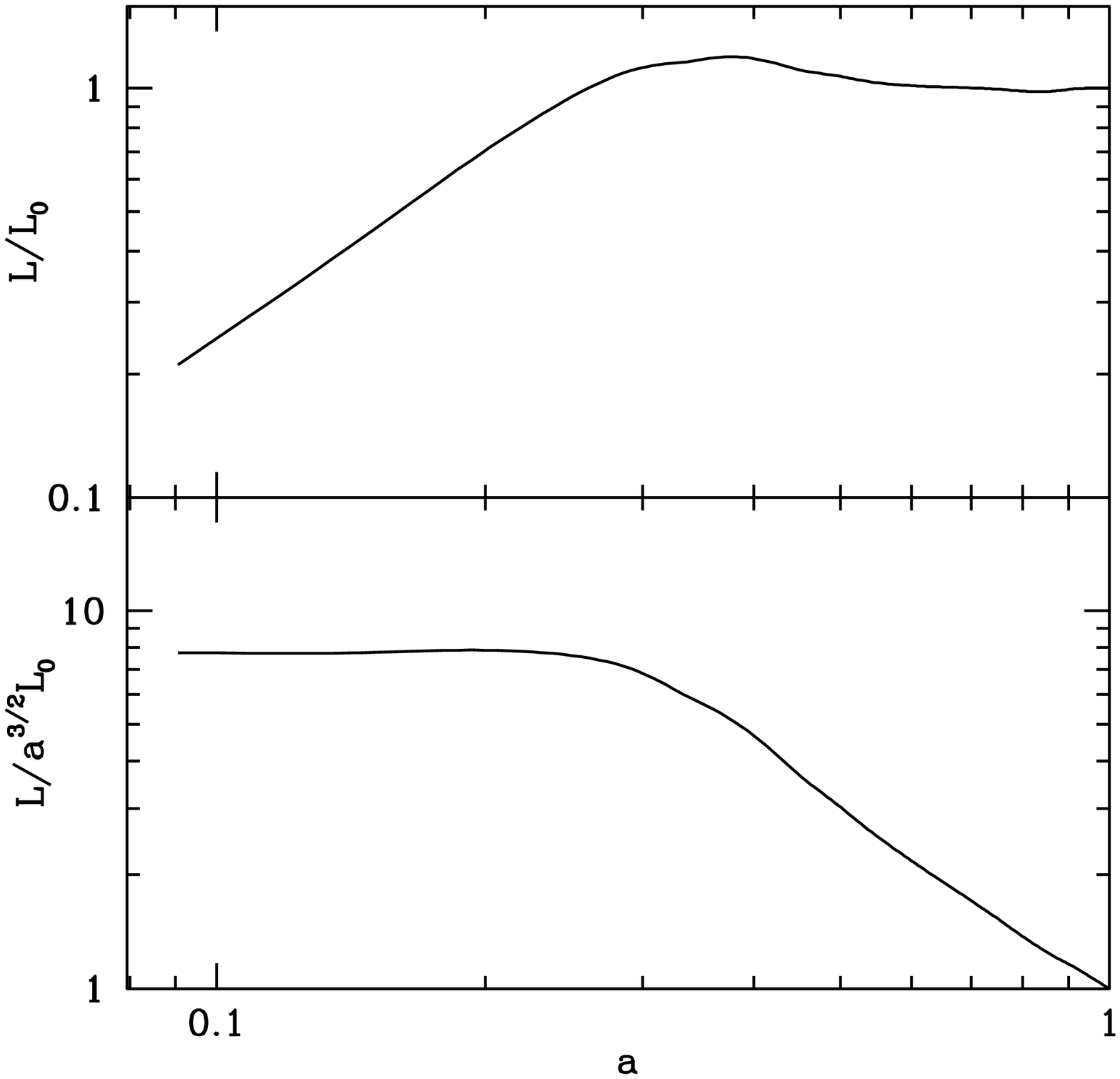}}\\
\subfloat[][]{\includegraphics[width=0.4\textwidth]{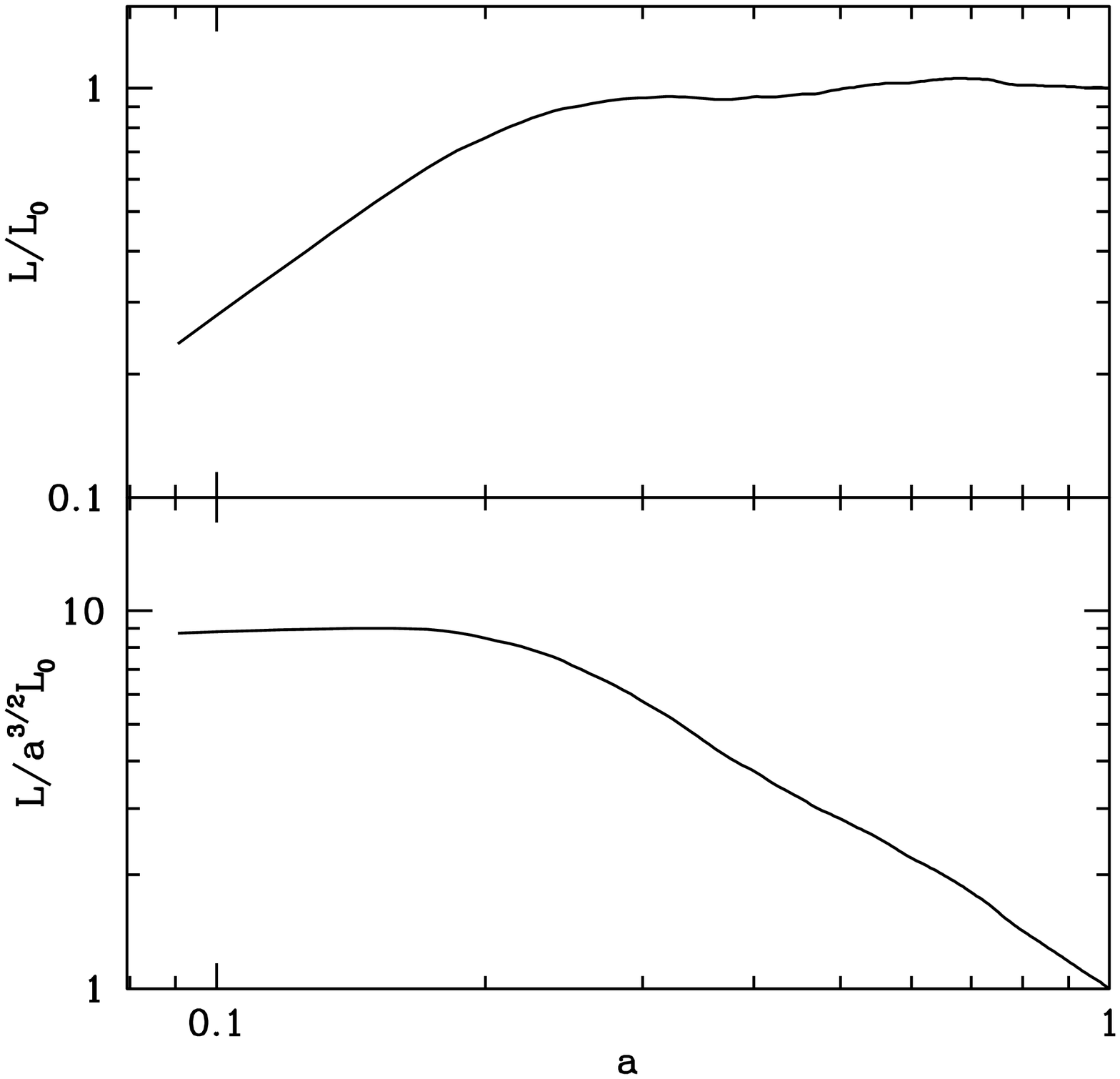}}
\subfloat[][]{\includegraphics[width=0.4\textwidth]{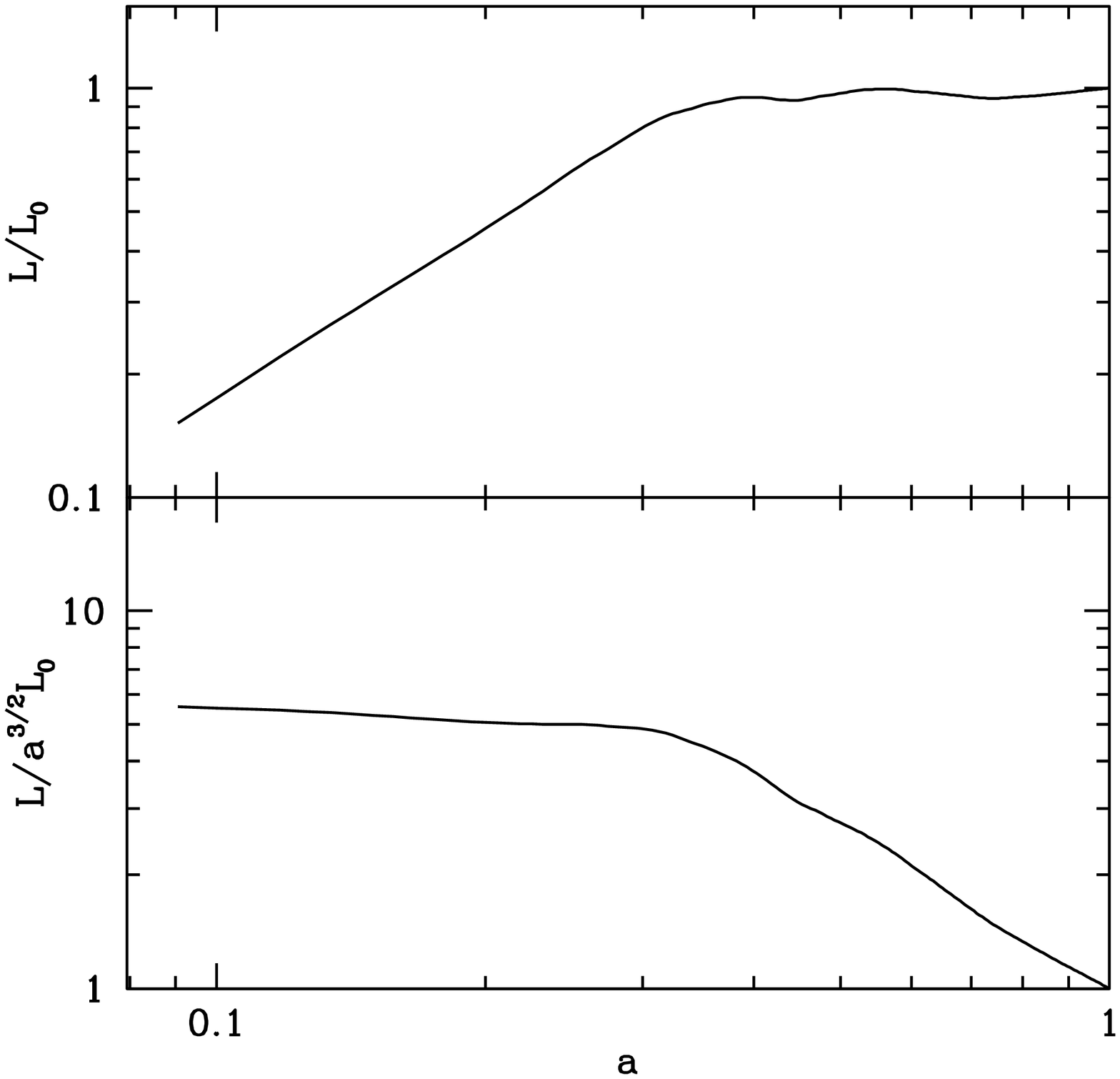}}
\caption{The evolution of the specific halo angular momentum, \textit{with respect to the centre of mass of the system}, of all particles that are within the virial radius of the halo at $z=0$ for halos (a) 239, (b) 258, (c) 277 and (d) 285. All halos follow a basic trend that agrees with the results of tidal torque theory \citep{White84,Zavala08}.}
\label{fig:zavala}
\end{figure*}

Before we describe the evolution of the angular momentum distribution in the simulations outlined in Sec. \ref{sec:sim}, we show the results of two simple checks.  First, to determine that any evolution in the angular momentum distribution is real and not a numerical artifact, we simulated a $10^{12}$ \Msun, spherically-symmetric halo with equilibrium initial conditions, which in the absence of a spurious numerical effect would maintain a static distribution of angular momenta.  These initial conditions were used in \citet{peter2010c} in the context of decaying dark matter\footnote{No decay term was included in this work, however.}.  The orbits of $10^6$ \Msunsp particles were integrated for 10 Gyr using PKDGrav \citep{Stadel01}.  We found no secular drift in the angular momentum distribution of particles in the halo, just as we expected for a spherically-symmetric halo in equilibrium. Further, we found that the angular momentum distribution in radial shells is invariant with time. Thus, we determined that the angular momentum evolution that we see in the cosmological case is not due to numerical effects.

As a further check, we compared the evolution of the total centre-of-mass angular momentum of our cosmological models with that found by other authors.  In Fig. \ref{fig:zavala}, we plot the specific halo angular momentum \textit{with respect to the centre of mass of the system} of all particles that are within the virial radius of the halo at $z=0$ (i.e. the Lagrangian region corresponding to the $z=0$ halo) for halos (a) 239, (b) 258, (c) 277 and (d) 285, as a function of the cosmological scale factor $a$. Specifically, in the top panel of each panel we show the specific angular momentum $L$, as defined above in the centre-of-mass frame, divided by its value $L_0$ at $z=0$. Tidal torque theory, as was first calculated by \citet{White84} and was observed in simulations by \citet{Zavala08}, implies that the specific angular momentum of a halo grows as $a^{3/2}$ until the halo virializes, at which point it becomes roughly constant. In the lower panels, we show the $L/L_0$ now divided by $a^{3/2}$ to explicitly remove this expected dependence. We find that in most of our halos  the angular momentum defined with respect to the centre of mass of the system behaves in the manner expected from tidal torque theory. Halo 239 also follows this trend but has significantly bumpier evolution than the other halos, possibly as a result of its continuously active merger history.

\subsection{Validity of the Adiabatic Approximation}  \label{sec:adiabatic}

To determine how well the prescription of adiabatic contraction is followed in simulations, we must first determine to what extent the potential of a halo evolves adiabatically. The adiabatic contraction model is only applicable if the typical time scale for order unity changes to the halo gravitational potential be long compared to the dynamical time of a typical particle in the halo. If this is not the case it would be an indication that the adiabatic contraction approximation may not be used\footnote{This does not, of course, address the issue of whether changes in the potential remain adiabatic once baryonic physics are included.}.

\begin{figure}
\centering
\includegraphics[width=0.47\textwidth]{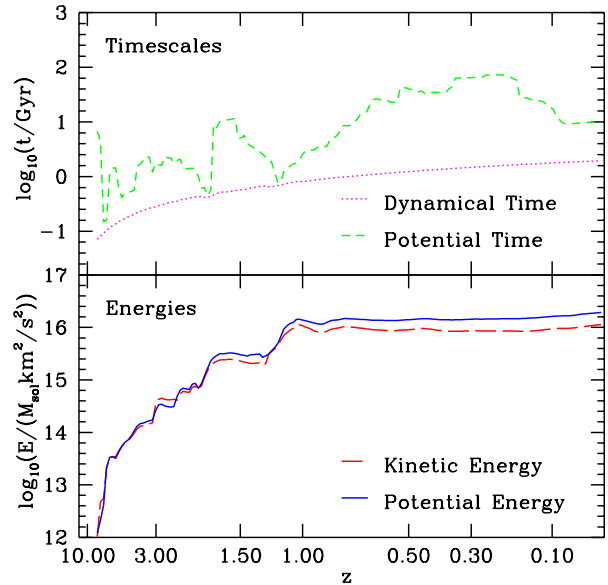}
\caption{Energies and time scales of all particles within the virial radius at each timestep. The halo shown is halo 239. The top panel shows the dynamical time of the entire halo vs. the time scale for changes to the gravitational potential, which has been smoothed over the dynamical time of the halo. The bottom panel shows the evolution of the potential and kinetic energy of all particles within the virial radius at each time step.}
\label{fig:tu}
\end{figure}

\begin{figure}
\centering
\includegraphics[width=0.47\textwidth]{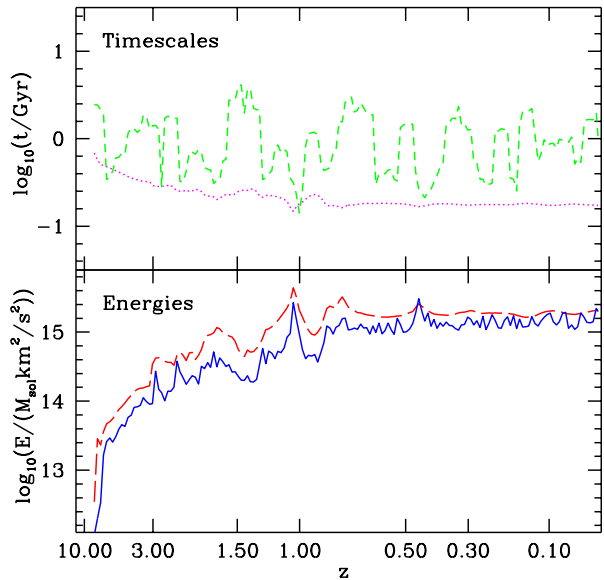}
\caption{This figure shows the same quantities as Fig.~\ref{fig:tu}, but for only those halo particles that are at each time step within a physical radius $0.1 r_{vir}(z=0)$ of the centre of the halo.}
\label{fig:tuinner}
\end{figure}

To answer this question, we examined the energies and time scales of all particles within the virial radius of each halo.  We show these quantities for halo 239 in Figs. \ref{fig:tu} and \ref{fig:tuinner}.  The top panel in Fig. \ref{fig:tu} shows the dynamical time of the entire halo, defined as the virial radius divided by the virial velocity, with the time scale for the potential of the entire halo to change by order unity. The potential change time scale is defined as $t_{\rm pot} = \left|\Phi/\left({\rm d}\Phi/{\rm d}t\right)\right|$, and has been smoothed on the dynamical time of the halo. It can be seen that, in general, the dynamical time is at least an order of magnitude smaller than the time scale for the gravitational potential to change by order unity, implying that the adiabatic approximation is reasonably accurate in these regions. However, when a halo experiences a merger, the potential change time scale is reduced and the evolution of the potential is non-adiabatic. Such mergers occur here at redshifts $2.2$ and $1.2$, with mass ratios of 1.1 and 1.9 respectively. The lower panel of Fig. \ref{fig:tu} shows the evolution of the potential and kinetic energy of all particles within the virial radius at each timestep, and shows that the halo reaches a stable quasi-equilibrium around a redshift of $z=1$. Overall, we see that the halo is roughly adiabatically evolving, particularly after this redshift.

As galaxies generally form in the inner regions of their host halos, we would also like to determine whether the adiabatic approximation is valid in the innermost regions of halos. This can be seen in Fig. \ref{fig:tuinner}, which shows the same quantities as Fig.~\ref{fig:tu} but for only those halo particles that are within the Eulerian-selected region $r(z) < 0.1\text{ }r_{\rm vir}(z=0)$. The potential change time scale of the inner halo is generally much closer to the dynamical time of this region, implying that the adiabatic approximation is less valid for the inner halo. We can see from the evolution of the kinetic and potential energies that the inner region reaches a state of stable quasi-equilibrium around $z=1$, similar to the outer regions. However, we notice that halo 239 has a more rapid change in angular momentum than the other halos (see figure \ref{fig:zavala}. Thus, we would expect the other halos to have larger angular momentum change timescales, and for the adiabatic approximation to be more correct in those cases.

These figures show that the approximation that the halo is evolving adiabatically is roughly accurate for the halos studied herein, although the dynamical time is often only an order of magnitude or less smaller than the potential change time scale. The entire halo reaches a quasi-equilibrium by around $z=1$ here, while halos with more quiescent merger histories reach an equilibrium earlier. However, in the inner regions of the halo where a galaxy might form, the adiabatic approximation is not as good, with the potential time scale usually above but frequently close to the dynamical time scale. Thus we must apply the assumption of adiabaticity with caution in the inner regions of the halo, and to some extent even in the outer regions, especially as work in the context of decaying dark matter has shown that AC works well only if the time scale for changes in the potential is much longer than the dynamical time \citep{peter2010c}.
 
\subsection{Angular Momentum Evolution}  \label{sec:Levol}

We next examine the extent to which angular momentum is conserved in our dark-matter-only halos. Here, we consider only angular momentum with respect to the centre of the halo, and not the centre of mass  \citep[as in, e.g., the comparison with][at the beginning of this section]{Zavala08}. The centre of the halo is determined by AHF, which uses an adaptive mesh to search for overdensities in the simulation. Once it identifies a halo, it defines the centre of the halo as the centre of mass of the particles on the highest refinement grid. We show the average particle angular momentum as it evolves over the course of the simulation for various groups of particles, chosen to emphasize particular physical characteristics of the system. In all cases, we bin particles radially so that the innermost ten percent of the halo particles are placed into bin 0, the next ten percent into bin 1, and so forth.

\begin{figure*}
\centering
\subfloat[][]{\includegraphics[width=0.49\textwidth]{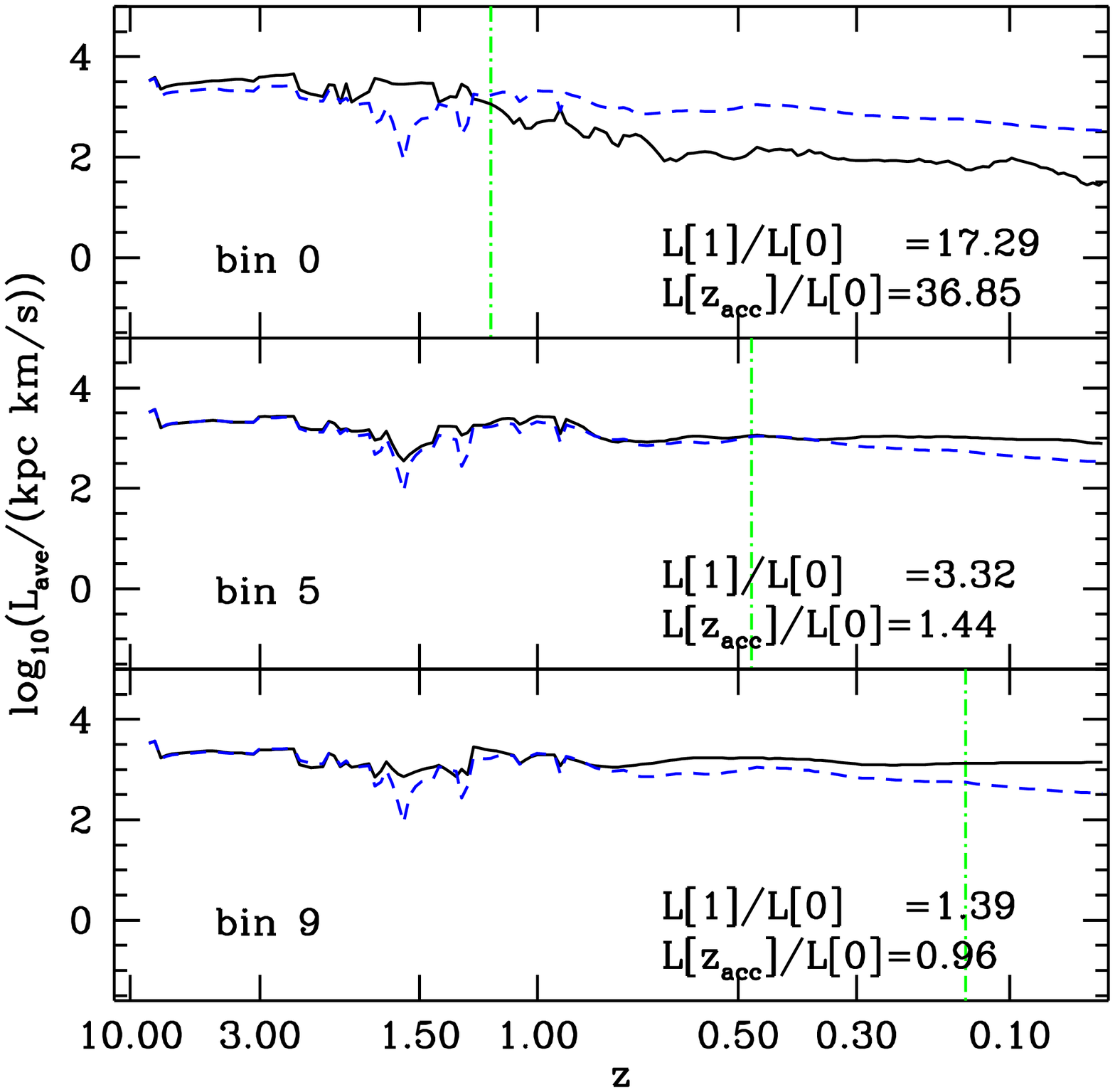}}
\subfloat[][]{\includegraphics[width=0.49\textwidth]{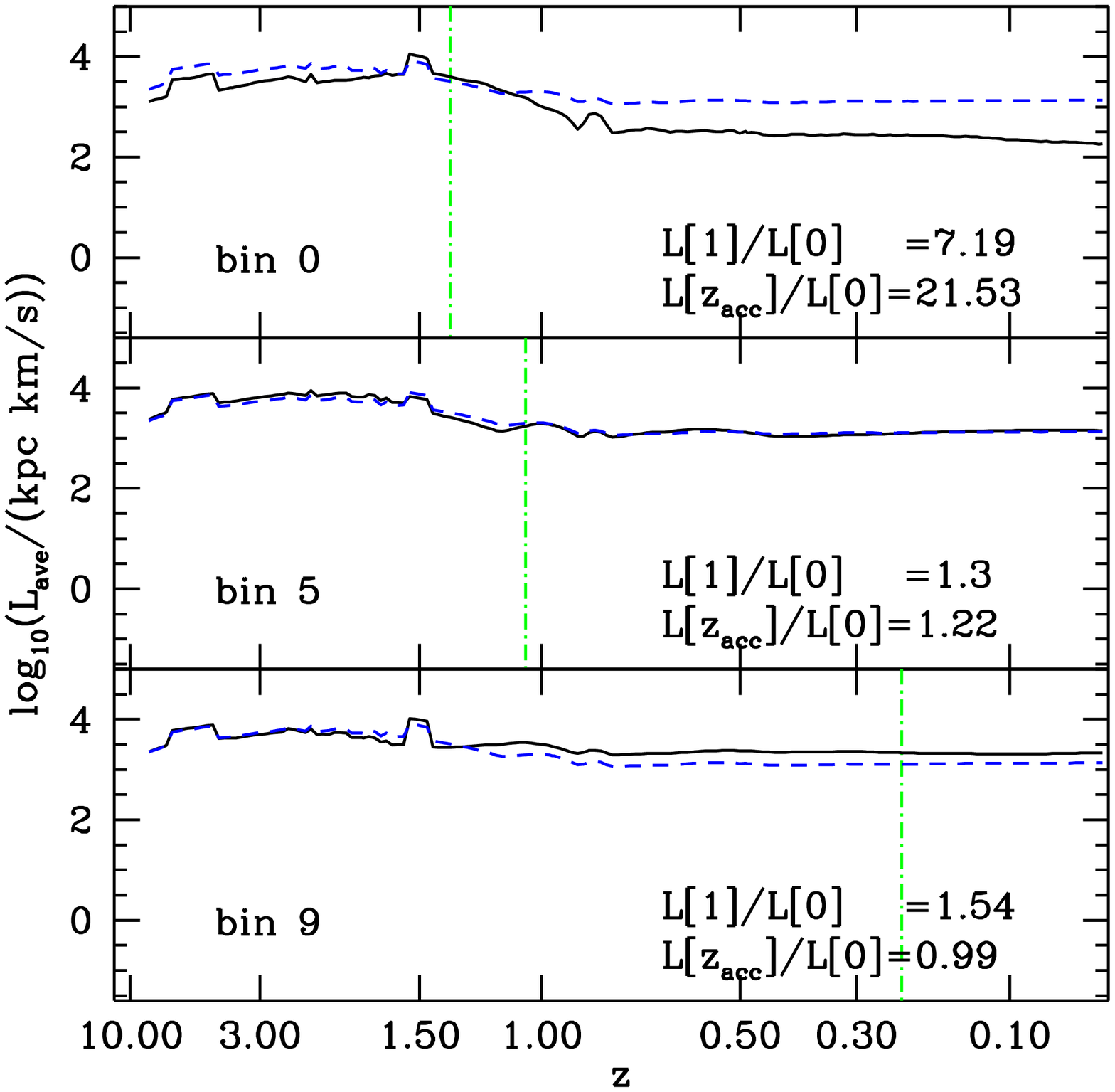}}\\
\subfloat[][]{\includegraphics[width=0.49\textwidth]{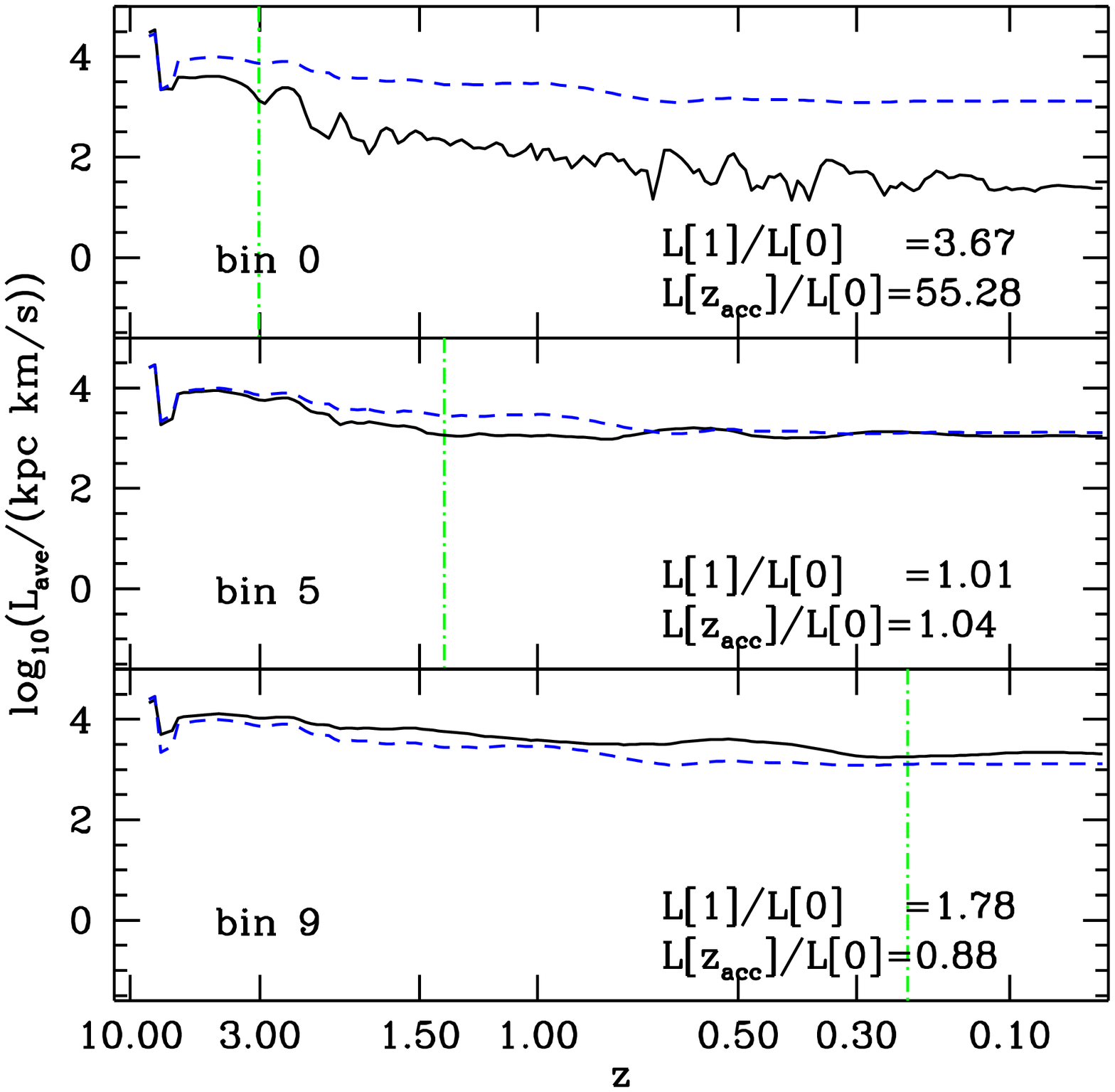}}
\subfloat[][]{\includegraphics[width=0.49\textwidth]{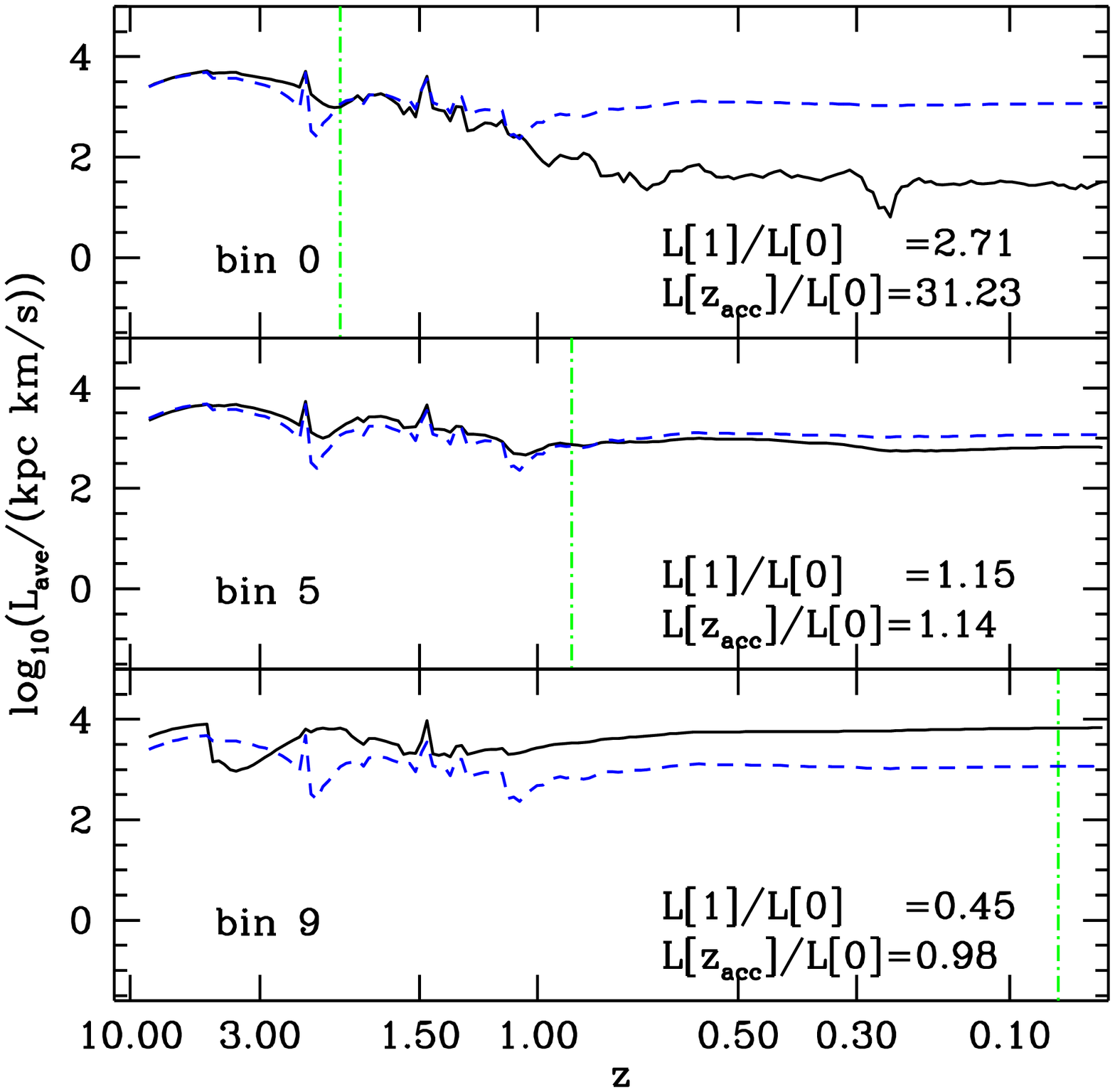}}
\caption{The evolution of the average angular momentum of particles chosen in a Lagrangian fashion to lie within the 0th, 5th and 9th of ten radial bins (solid black lines) compared to the evolution of the average angular momentum of all halo particles that are within the virial radius at $z=0$ (blue dashed lines), for halos (a) 239, (b) 258, (c) 277 and (d) 285. The angular momenta in these plots were vector-averaged, and we show the redshift at which each bin is accreted onto the halo (green dot-dashed vertical lines). In each panel is given the ratios of the angular momentum at $z=1$ and at the redshift of their accretion, $z_{\rm acc}$, onto the halo to the value at $z=0$.}
\label{fig:Ls}
\vspace{200pt}
\end{figure*}

\begin{figure*}
\centering
\subfloat[][]{\includegraphics[width=0.49\textwidth]{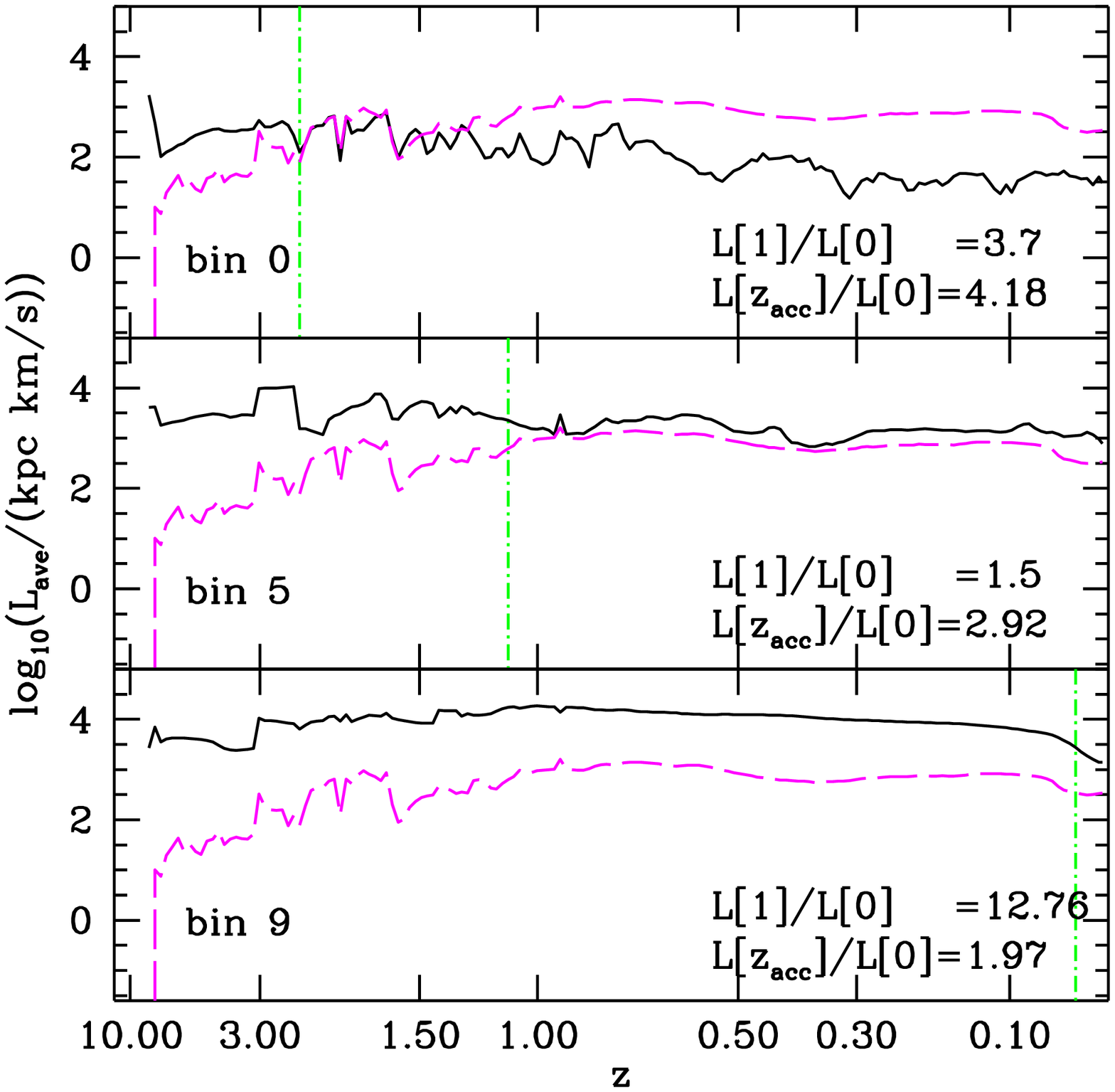}}
\subfloat[][]{\includegraphics[width=0.49\textwidth]{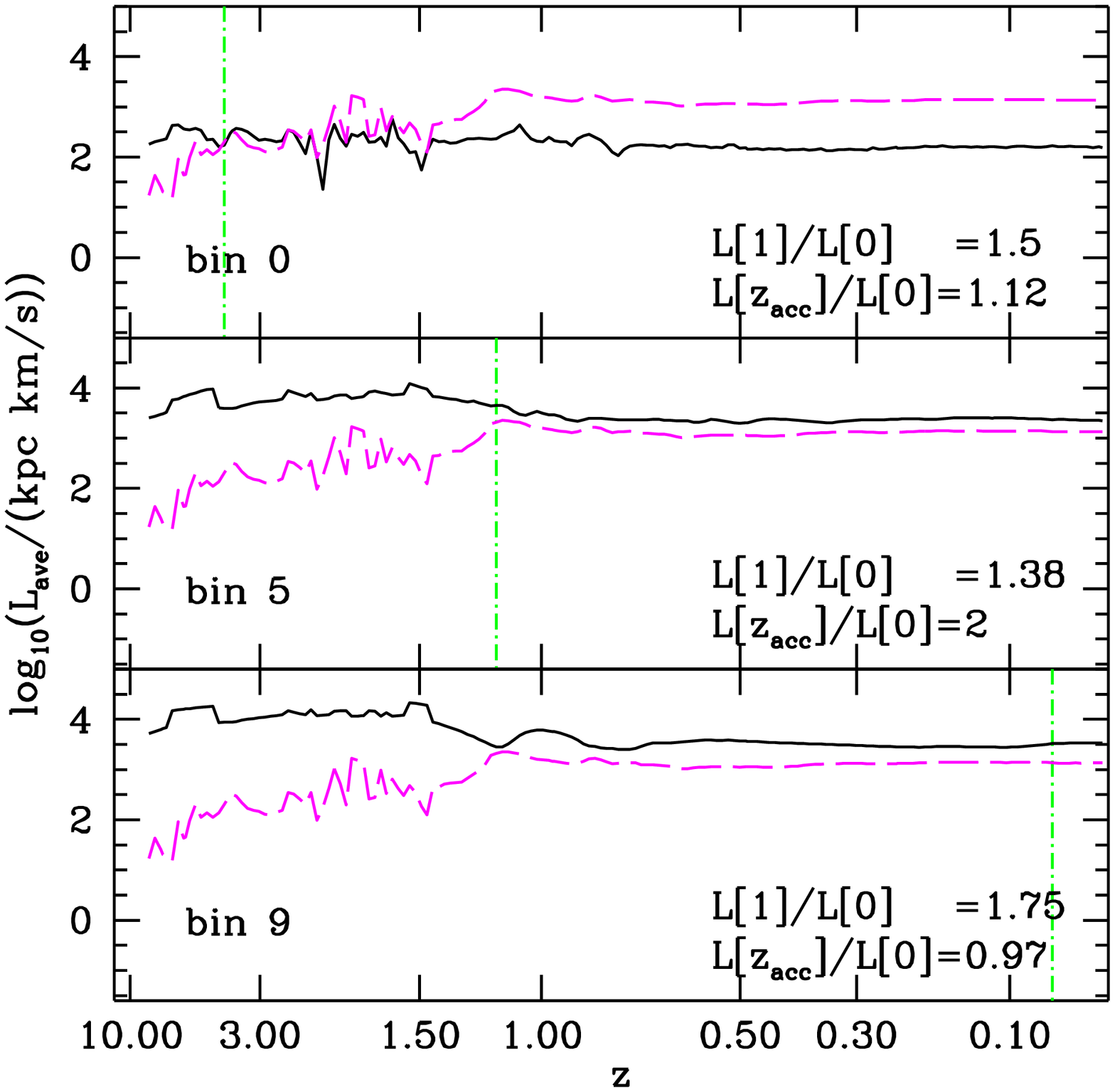}}\\
\subfloat[][]{\includegraphics[width=0.49\textwidth]{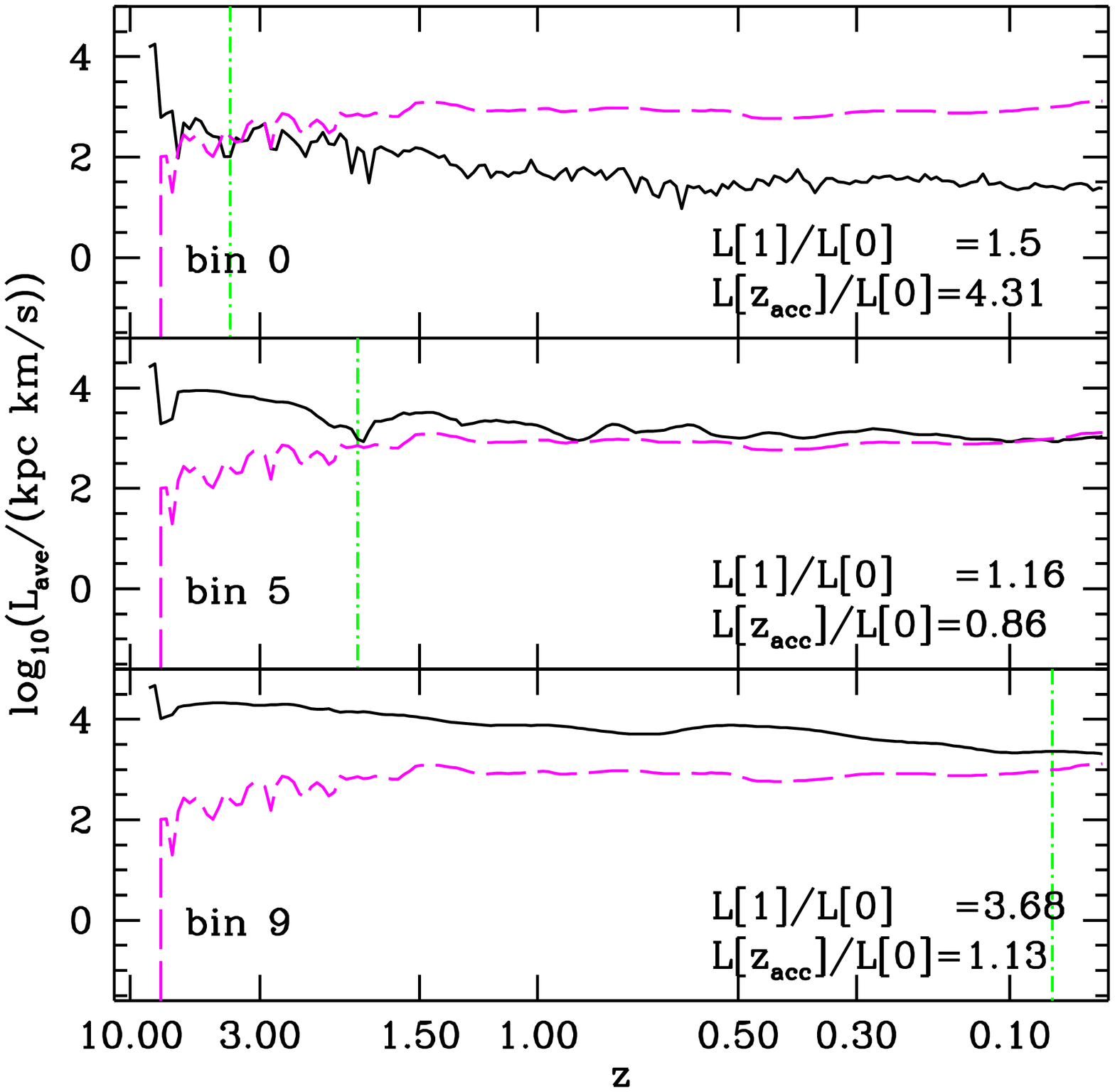}}
\subfloat[][]{\includegraphics[width=0.49\textwidth]{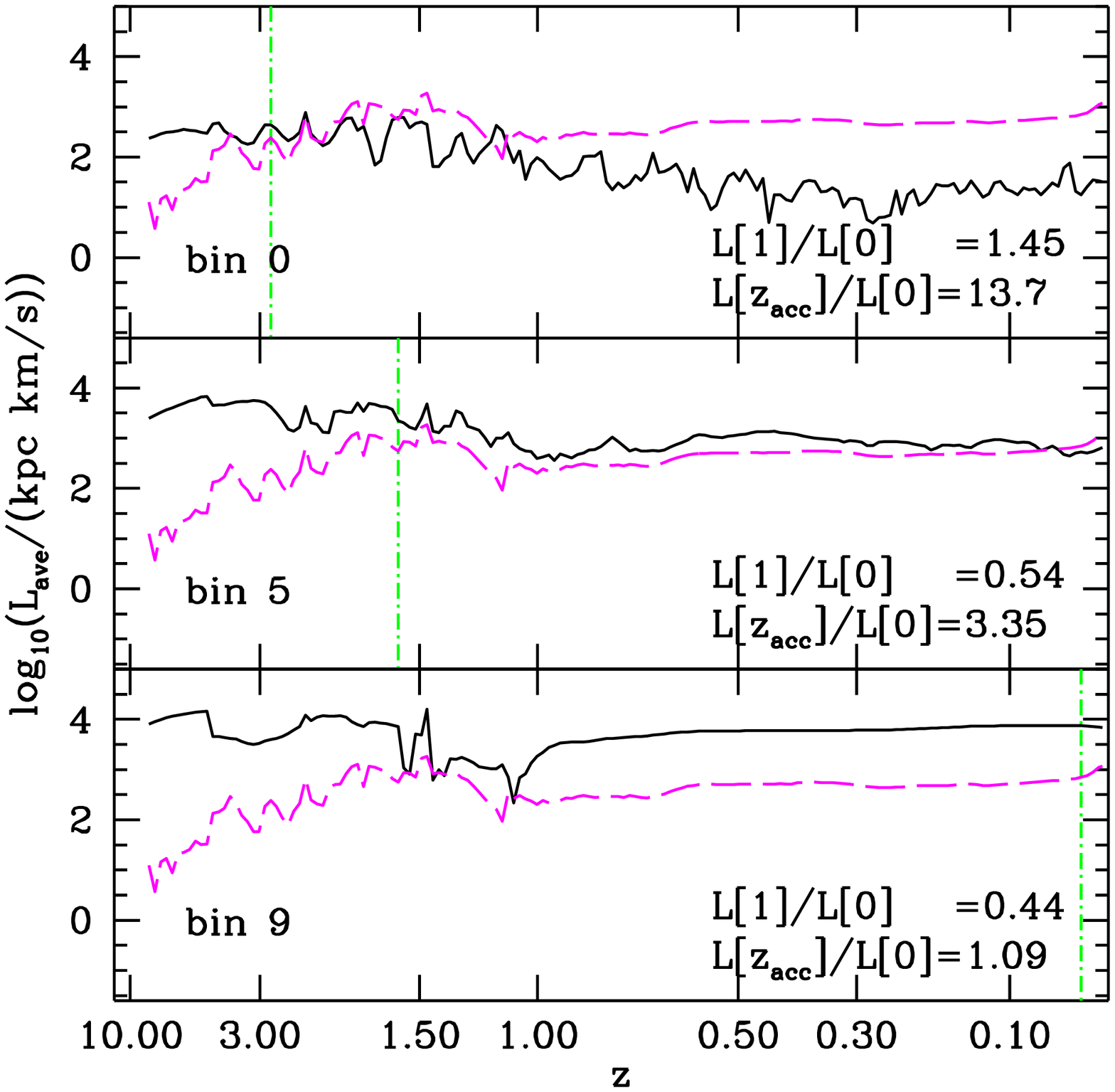}}
\caption{Similar to Fig. \ref{fig:Ls}, except that the particles here are binned radially in an Eulerian fashion. The magenta long-dashed line shows the specific angular momentum of all halo particles within the virial radius at the given redshift.}
\label{fig:Le}
\vspace{200pt}
\end{figure*}

We present this information in Figs. \ref{fig:Ls} and \ref{fig:Le}. In Fig. \ref{fig:Ls} we show the evolution of the vector-averaged angular momentum of particles chosen to lie within radial bins interior to the virial radius at $z=0$ (Lagrangian radial selection, solid black lines), compared to the evolution of the vector-averaged angular momentum of all particles within the virial radius at $z=0$ (blue dashed lines), for halos (a) 239, (b) 258, (c) 277 and (d) 285. The angular momenta in these plots were vector-added and then divided by the particle number to obtain the average, i.e., the specific angular momentum.  Particles were placed into 10 radial bins at $z=0$, and we show the 0th (innermost with respect to the halo centre), 5th (middle) and 9th (outermost) bins in the top, middle and bottom panels of each plot. We also show the redshift at which the average radius of particles in this radial bin passes within the virial radius of the halo (green dot-dashed vertical lines), and the ratios of the angular momentum at $z=1$ and at the redshift of their accretion onto the halo to illustrate quantitatively the extent of angular momentum loss. In this plot, we can see that generally, the particles that end up in the inner regions of these halos at $z=0$ lose a larger fraction of their angular momentum than those that end up in the outer regions, and the angular momentum of these $z=0$ inner particles also fluctuates more rapidly. The blue dashed line, which shows the evolution of the average angular momentum of all particles within the virial radius at $z=0$, shows the average behaviour of the particles in all bins, and, like the binned behaviour, always decreases.

As mentioned earlier, the angular momentum of particles selected based on their Eulerian radius is an interesting compliment to the Lagrangian case as it shows how those particles that are in a given radial bin at each timestep evolve in angular momentum, and thus more clearly illustrates the evolution of the radial angular momentum profile of the halo. By contrast, the angular momentum evolution of radially Lagrangian-selected particles shows how the angular momentum of those particles which end up in certain radial bins have changed over time. 

The quantities shown in Fig. \ref{fig:Le} are similar to those in Fig. \ref{fig:Ls}, except that the particles here (specifically, all particles  that are within the virial radius at $z=0$) were binned radially at each time step in an Eulerian fashion, so that the ``bin 0'' line shows the specific angular momentum of the ten percent of particles that are closest to the centre of the halo at each time step. The magenta long-dashed line shows the specific angular momentum of all particles that are within the virial radius at each time step. This quantity goes to zero at the beginning of the simulation as the virial radius is zero at the first time step and increases from there. This line tends to increase because the virial radius increases with time. This is different from the binned values which include all particles that are within the virial radius at $z=0$ and thus always contain a nonzero number of particles. All of the angular momenta in this plot were vector added as in the previous figure. 

In Fig.~\ref{fig:Le}, there is no clear pattern as to which bin loses a larger fraction of its angular momentum, although the inner bin generally has less angular momentum. This shows that, in general, particles throughout the halo lose a comparable fraction of their angular momentum, but those in the inner regions at each timestep tend to have less angular momentum than those in the outer regions. However, we saw in Fig.~\ref{fig:Ls} that when the particles were binned in a Lagrangian fashion, the particles in the inner regions lost a larger fraction of their angular momentum as well as having lower angular momentum in general. Noting that both figures \ref{fig:Le} and \ref{fig:Ls} must and do converge to the same value at $z=0$, this implies that the particles that end up in the innermost bin at $z=0$ both begin with a larger angular momentum than those particles that are in the innermost bin at the beginning of the simulation, and subsequently lose this angular momentum at a faster rate than those particles that are at each timestep at the centre of the halo. This behaviour is to be expected, as the particles at the centre of the halo at each timestep will tend to have a low angular momentum, and it is reasonable to expect that this value will tend to drop rather slowly since we are choosing new, low-angular-momentum sets of particles at each timestep in the innermost bin.


We also examined the evolution of the angular momentum in radial shells when averaged using the magnitudes of particle angular momenta rather than vector-averaging. We have omitted plots of the magnitude-averaged cases as they are similar to Figs. \ref{fig:Ls} and \ref{fig:Le}, and have instead included the relevant information about their evolution in table \ref{tab:mag}. As we noted in section \ref{sec:subsets}, magnitude-averaged angular momenta tell us about the evolution of particle orbits, thus providing complementary information to the vector-averaged angular momenta that tell us about the evolution of the spin of radial shells. Further, it is the magnitude of angular momentum which is most relevant for the distribution function and adiabatic invariance.

The magnitude-averaged angular momenta evolve much more smoothly than the vector-averaged ones, indicating that cancellation between particles in the vector-averaged case causes more variability than in the magnitude-averaged case. We show in table \ref{tab:mag} the values of $L[1]/L[0]$ and $L[z_{acc}]/L[0]$ for these cases. The particles that end up in the inner regions tend to have a higher angular momentum to start with than those particles that are in the innermost bin at the beginning of the simulation, and subsequently tend to lose this angular momentum more rapidly, though the effect is not as pronounced as in Fig. \ref{fig:Ls}. The more pronounced nature of this effect in Fig. \ref{fig:Ls} indicates that the inner particles also tend to have their directions scrambled (i.e. are more thoroughly virialized) more than those in outer regions.

The mean magnitude-averaged angular momentum in each radial bin also decreases from the time of accretion to the present, while the vector-averaged angular momenta in Fig.~\ref{fig:Le} in general decrease more since their accretion and have lower values. In both cases, the inner regions lose a larger fraction of their angular momentum than the outer regions. It should be noted that the trend of inner particles losing more angular momentum than outer particles was also found by \citet{Zavala08}. The generally larger decrease over time in the vector-averaged angular momenta implies that the scrambling, or decoherence, of angular momentum direction in the halo tends to increase over time.

Thus, we have found that those particles that end up in the inner regions of halos lose a larger fraction of their angular momentum than other halo particles, and that all of the particles in the halo tend to have the direction of their angular momenta scrambled progressively more over time. Further, we see that those particles that end up in the inner regions of the halo at $z=0$ tend to start out with a higher angular momentum than those particles that are in the innermost bin at each timestep, and that they tend to lose this angular momentum more quickly.

\begin{table*}
\centering
\caption{ Ratios of the average binned angular momentum at z=1 and at accretion of the radial bin to the value of the bin's average angular momentum at z=0 for both Lagrangian and Eulerian selected radial bins, differing from figures \ref{fig:Ls} and \ref{fig:Le} in that the angular momenta are magnitude-averaged rather than vector-averaged.\label{tab:mag} }
\begin{tabular}{ | l | r r | r r | r r | r r | r r | r r | }
\hline
\hline
&\multicolumn{6}{ | c | }{Lagrangian} & \multicolumn{6}{ | c | }{Eulerian} \\
\hline
&\multicolumn{2}{ | c | }{Bin 0} & \multicolumn{2}{ | c | }{Bin 5} & \multicolumn{2}{ | c | }{Bin 9} & \multicolumn{2}{ | c | }{Bin 0} & \multicolumn{2}{ | c | }{Bin 5} & \multicolumn{2}{ | c | }{Bin 9}\\
&L$_1$/L$_0$ & L$_{acc}$/L$_0$ & L$_1$/L$_0$ & L$_{acc}$/L$_0$ & L$_1$/L$_0$ & L$_{acc}$/L$_0$ & L$_1$/L$_0$ & L$_{acc}$/L$_0$ & L$_1$/L$_0$ & L$_{acc}$/L$_0$ & L$_1$/L$_0$ & L$_{acc}$/L$_0$ \\
\hline
H239 & 2.24 & 3.45 & 1.44 & 1.27 & 1.41 & 1.07 & 1.29 & 1.22 & 1.2 & 1.2 & 2.52 & 1.05 \\
H258 & 2.42 & 5.96 & 1.21 & 1.23 & 1.3 & 1.03 & 1.31 & 1.31 & 1.08 & 1.29 & 1.81 & 0.99 \\
H277 & 1.63 & 3.31 & 1.31 & 1.45 & 1.38 & 1.06 & 1.18 & 1.64 & 1.19 & 1.08 & 1.68 & 1.01 \\
H285 & 1.22 & 3.55 & 1.11 & 1.1 & 1.05 & 0.99 & 0.95 & 1.36 & 1.04 & 1.12 & 1.28 & 1.02 \\
\hline
\end{tabular}
\end{table*}

\subsection{Distribution of Angular Momentum}  \label{sec:Distrib}

\begin{figure*}
\centering
\subfloat[][]{\includegraphics[width=0.49\textwidth]{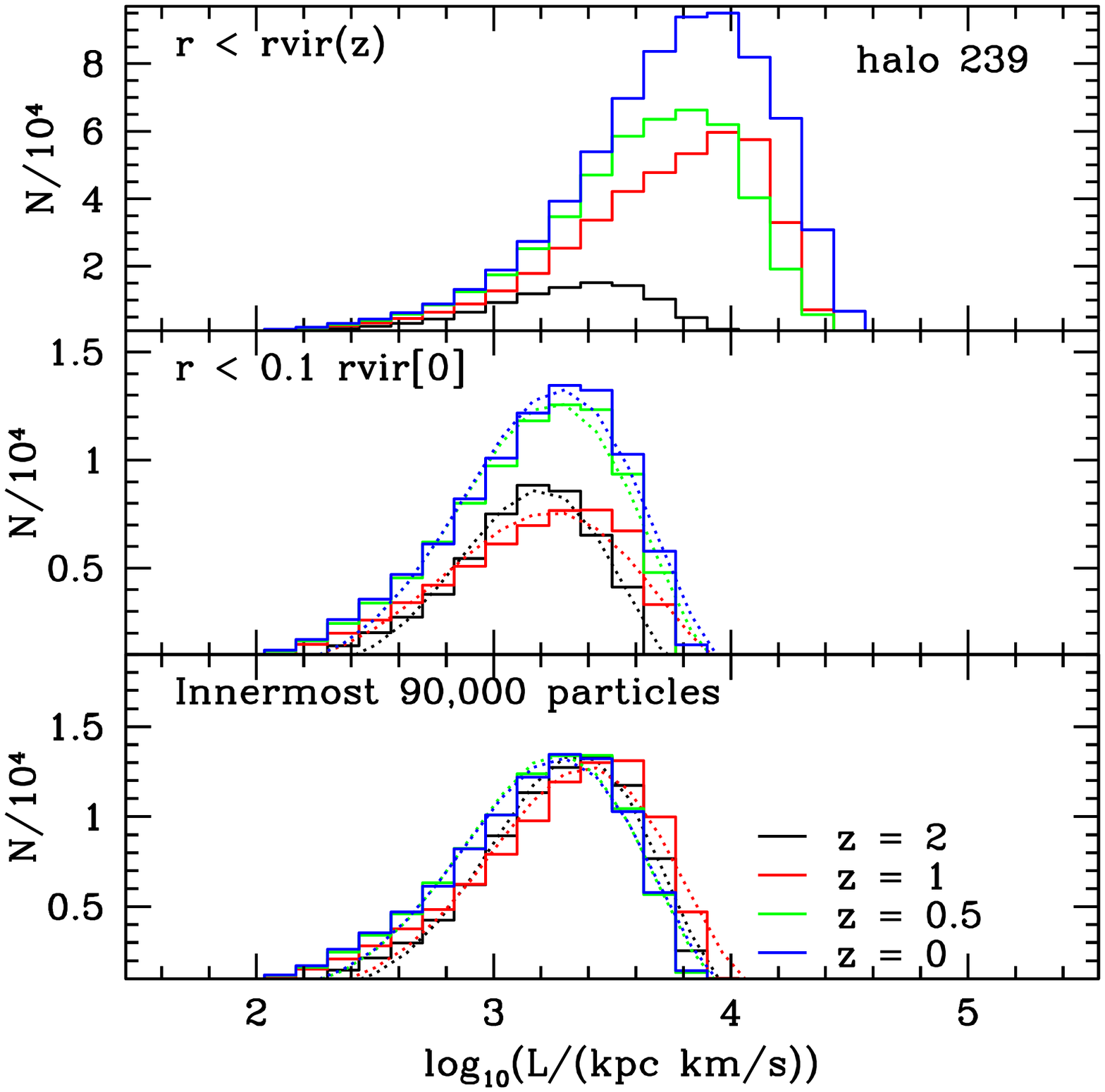}}
\subfloat[][]{\includegraphics[width=0.49\textwidth]{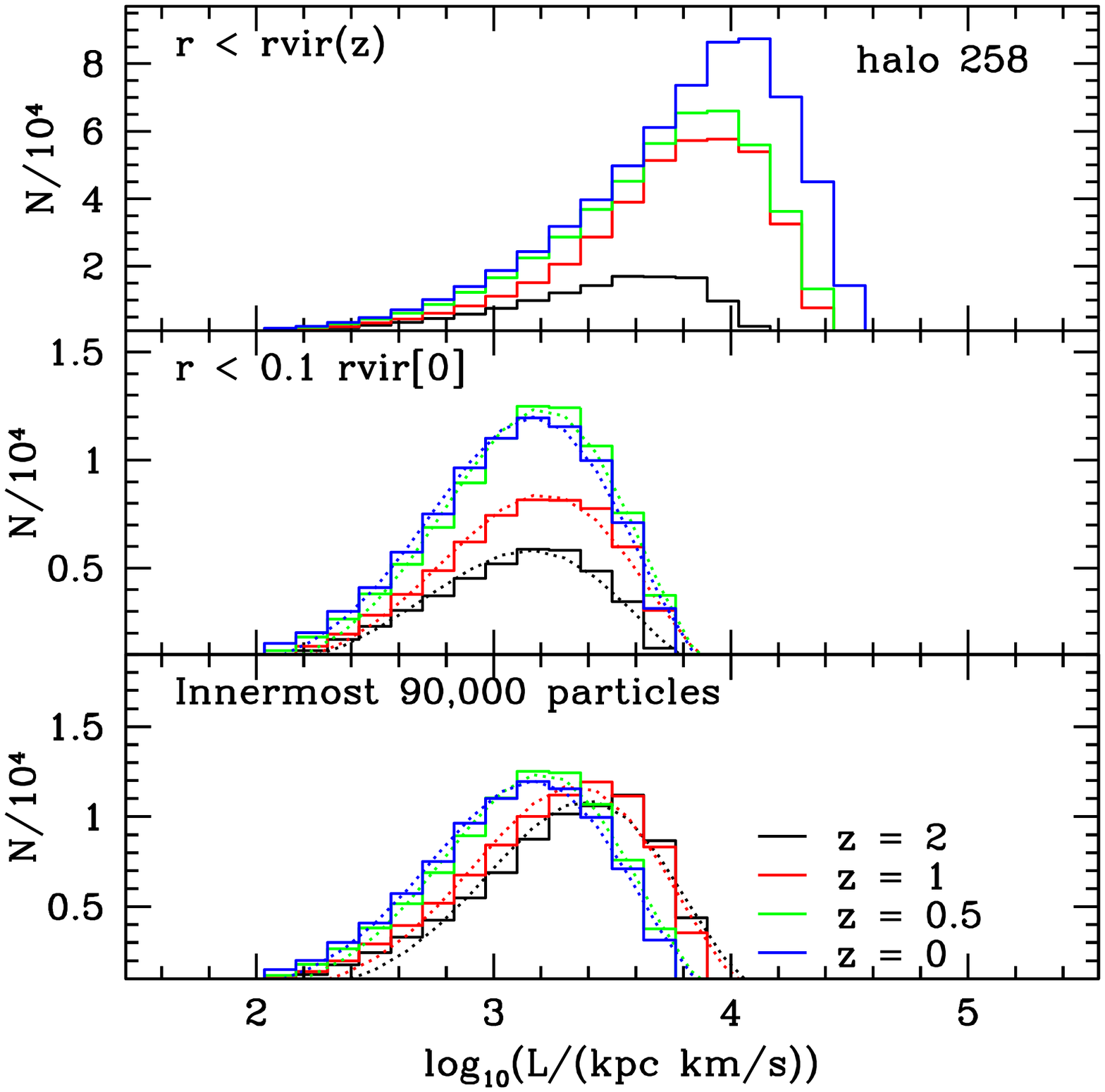}}\\
\subfloat[][]{\includegraphics[width=0.49\textwidth]{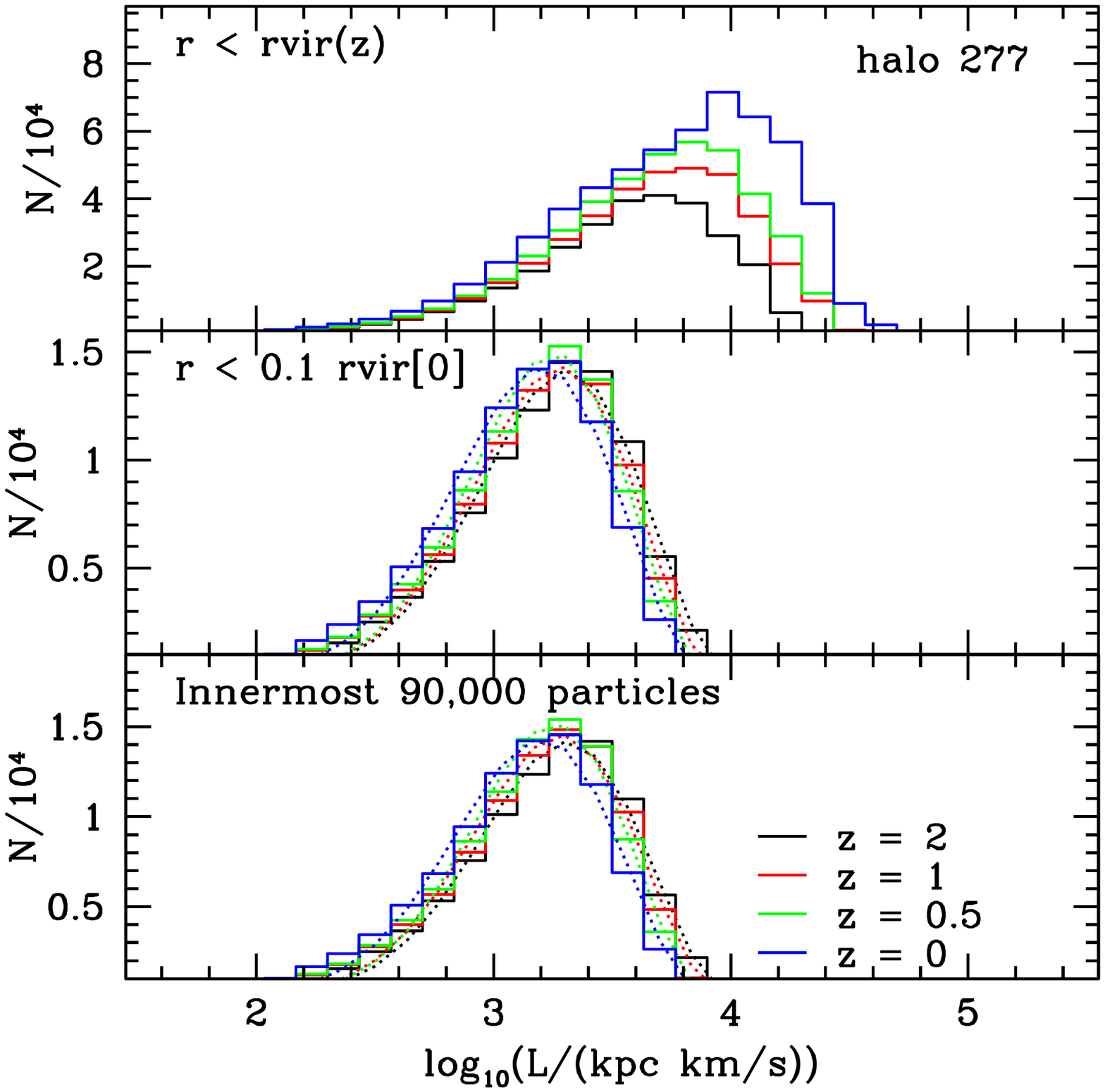}}
\subfloat[][]{\includegraphics[width=0.49\textwidth]{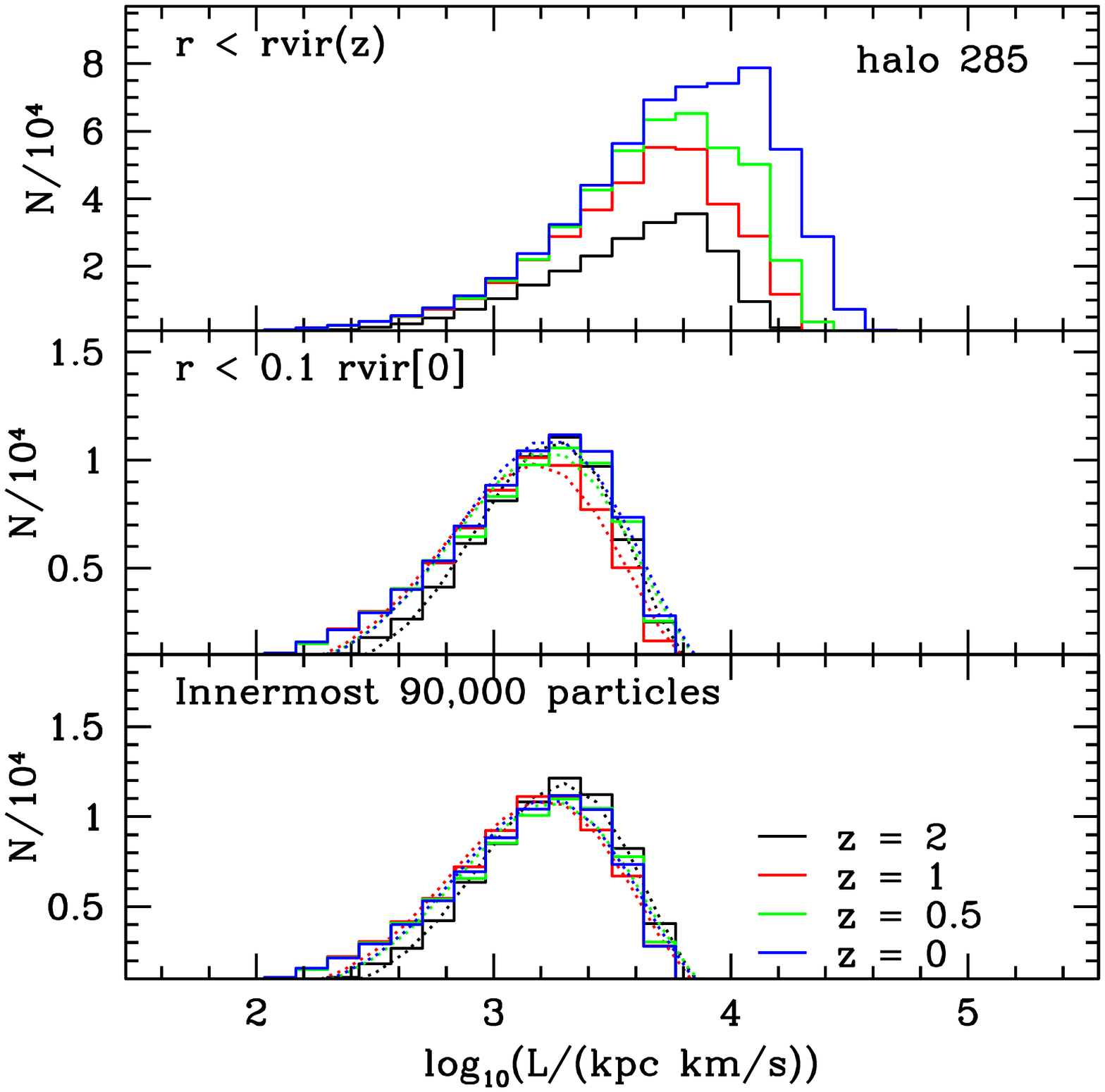}}
\caption{The evolution of the angular momentum of halo dark-matter particles for halos (a) 239, (b) 258, (c) 277 and (d) 285. In each figure, the top panel shows the angular-momentum distribution of all particles within the virial radius selected in an Eulerian fashion. In the middle and bottom panels, we present the evolution of `inner' halo particles, also selected in an Eulerian fashion, chosen radially (middle) and by mass (bottom). We also show in these panels the fit $N = A\text{ }(\log_{10}(L/(\text{kpc km/s})) - \log_{10}(L_0))^2 \exp\left(-(\log_{10}(L/(\text{kpc km/s})) - \log_{10}(L_0))^2/2 a^2\right)$ as dotted lines.}
\label{fig:fits}
\vspace{200pt}
\end{figure*}

Next, we consider changes to the distribution of the magnitudes of the particle angular momenta, not just the average as shown in the last section. In Fig. \ref{fig:fits}, we show the distributions of angular momenta of all halo particles and inner halo particles for all four halos. The top panel shows the angular momentum distribution of all particles within the virial radius chosen in an Eulerian fashion. Here, the number of particles, as well as their mean angular momentum, increases with time. In the middle and bottom panels, we show the evolution of `inner' halo particles chosen in an Eulerian fashion, defined as those particles within $0.1\text{ }r_{\rm vir}(z=0)$ (middle) and the innermost $90000$ particles corresponding to the innermost $10^{11} M_{\odot}$ (bottom). We also show in these panels the Boltzmann parametric fit $N = A\text{ }(\log_{10}(L/(\text{kpc km/s})) - \log_{10}(L_0))^2 \exp\left(-(\log_{10}(L/(\text{kpc km/s})) - \log_{10}(L_0))^2/2 a^2\right)$. The number of particles in the middle panel increases with time, while in the bottom panel it remains constant. 

The inner particles have a lower average angular momentum than all halo particles, as is to be expected. The evolution of the distribution of angular momenta of all halo particles is qualitatively similar in all four halos, with the average angular momentum of halo particles increasing with time as higher angular momentum particles are accreted.

Looking at the middle and lower panels, we can see that for halo 277 the angular momentum distribution stays roughly constant in the inner regions, with few extra particles entering $0.1\text{ }r_{\rm vir}$ at each time step. There is, however, a gradual decline in average angular momentum with time. This smooth evolution is related to the relatively quiescent evolution of halo 277. By contrast, we find that the innermost $10^{11} M_{\odot}$ of halo 258 have a significantly lower average angular momentum after $z=1$ than before, reflecting a major merger around this redshift. In general, all of these angular momentum profiles have a similar shape, which we find is fit well by the above fitting function.

In all the halos, though, there are changes of the angular momentum distribution with redshift. This implies that the angular momentum of a particle is strictly not an adiabatic invariant (or, perhaps, the time scale for changes is not sufficiently slow for the adiabatic approximation to hold). This means that assuming that the distribution function is constant when expressed in terms of the angular momentum during periods when a galaxy is growing is formally incorrect. In practice, this lack of precise conservation simply sets a limit on how accurate the AC approach can ever be.

\begin{figure*}
\centering
\subfloat[][]{\includegraphics[width=0.49\textwidth]{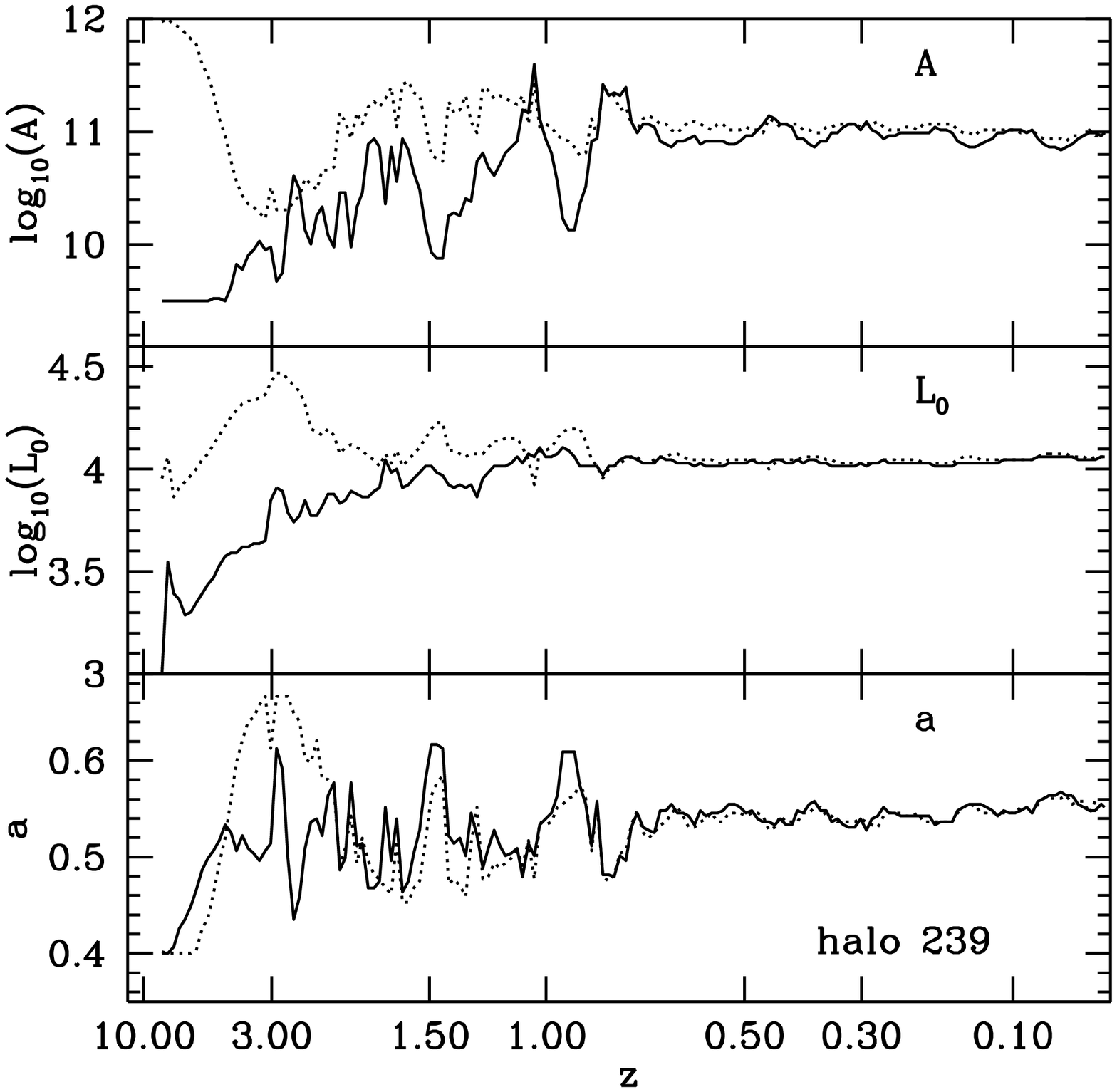}}
\subfloat[][]{\includegraphics[width=0.49\textwidth]{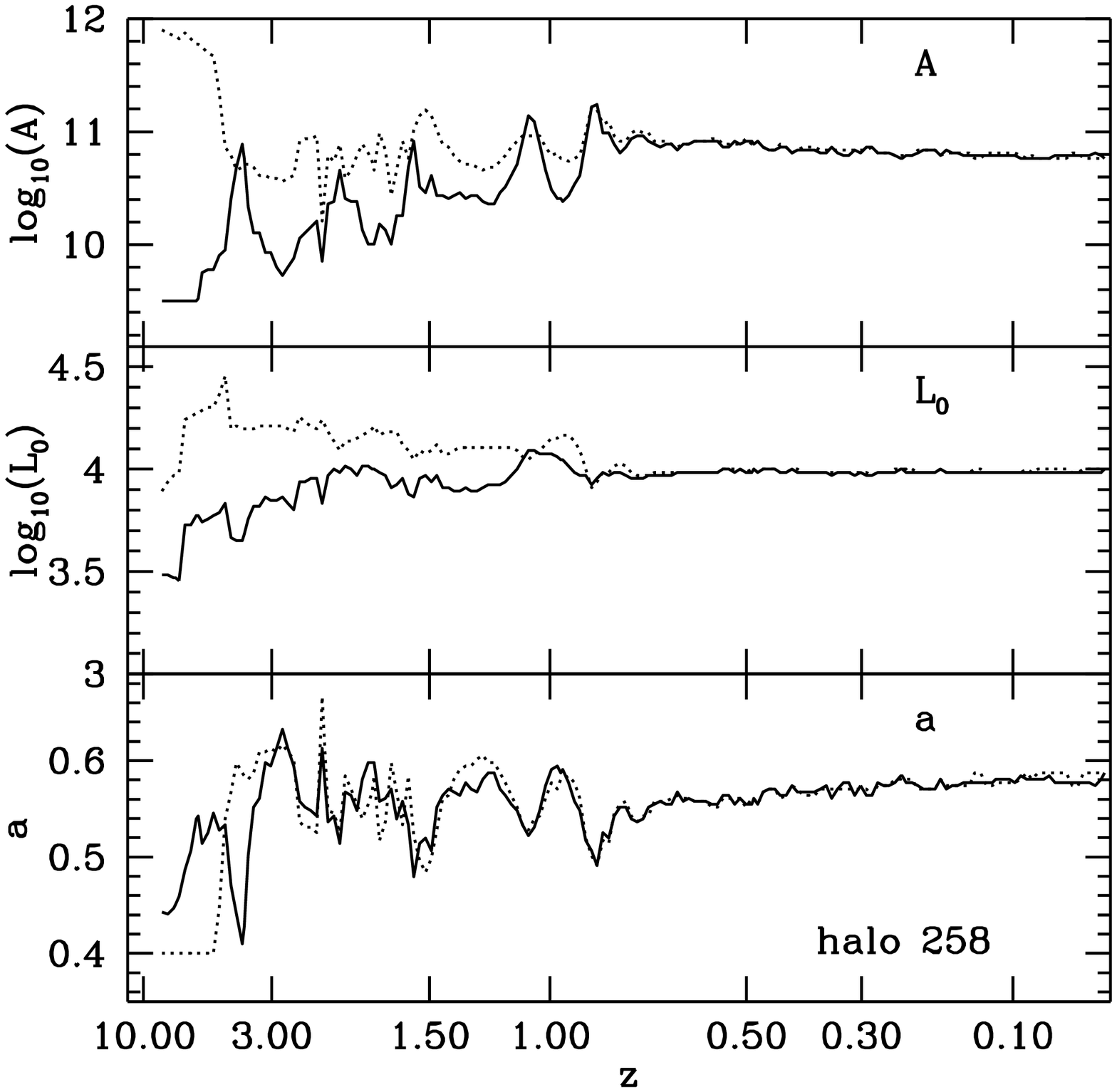}}\\
\subfloat[][]{\includegraphics[width=0.49\textwidth]{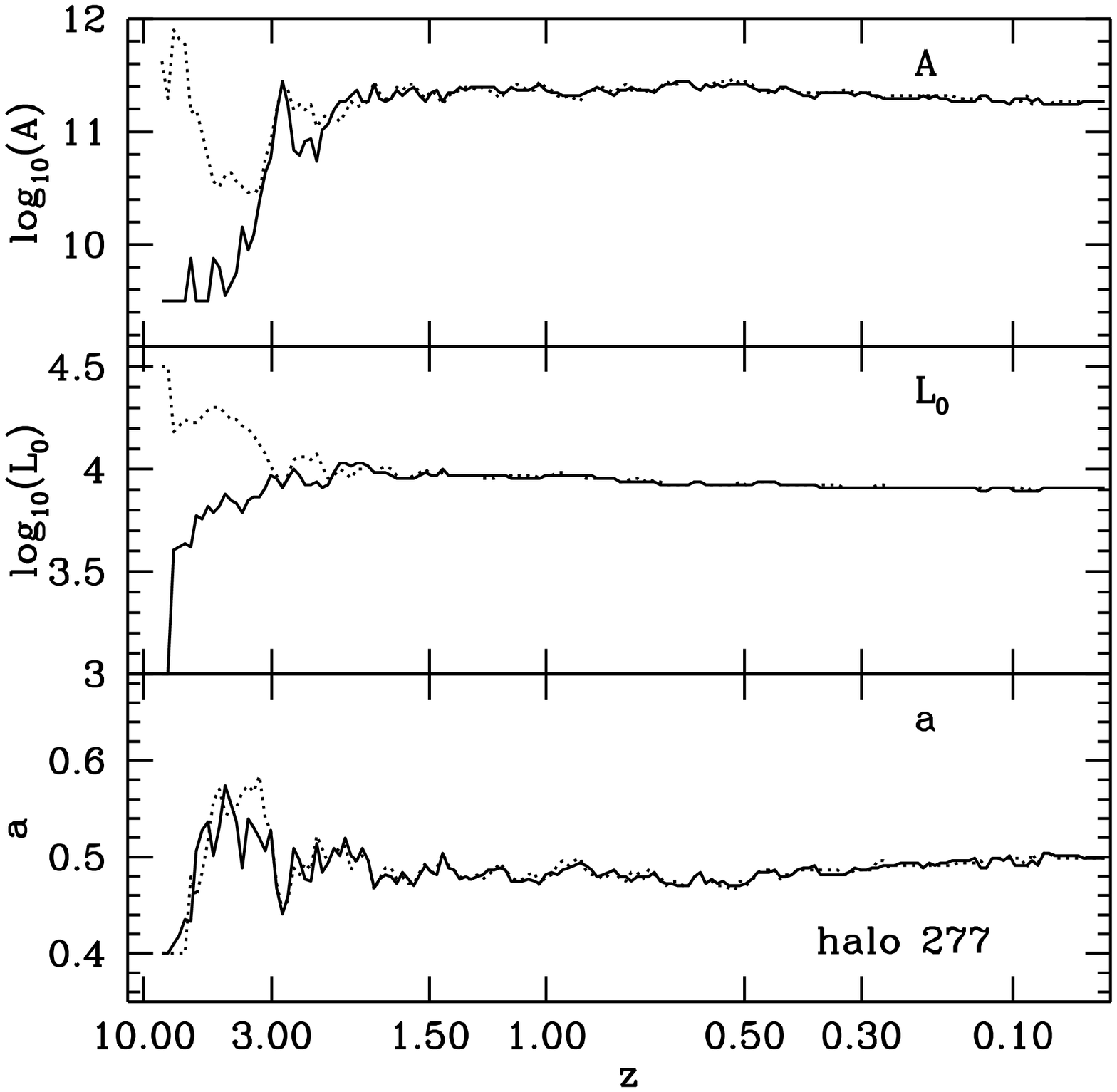}}
\subfloat[][]{\includegraphics[width=0.49\textwidth]{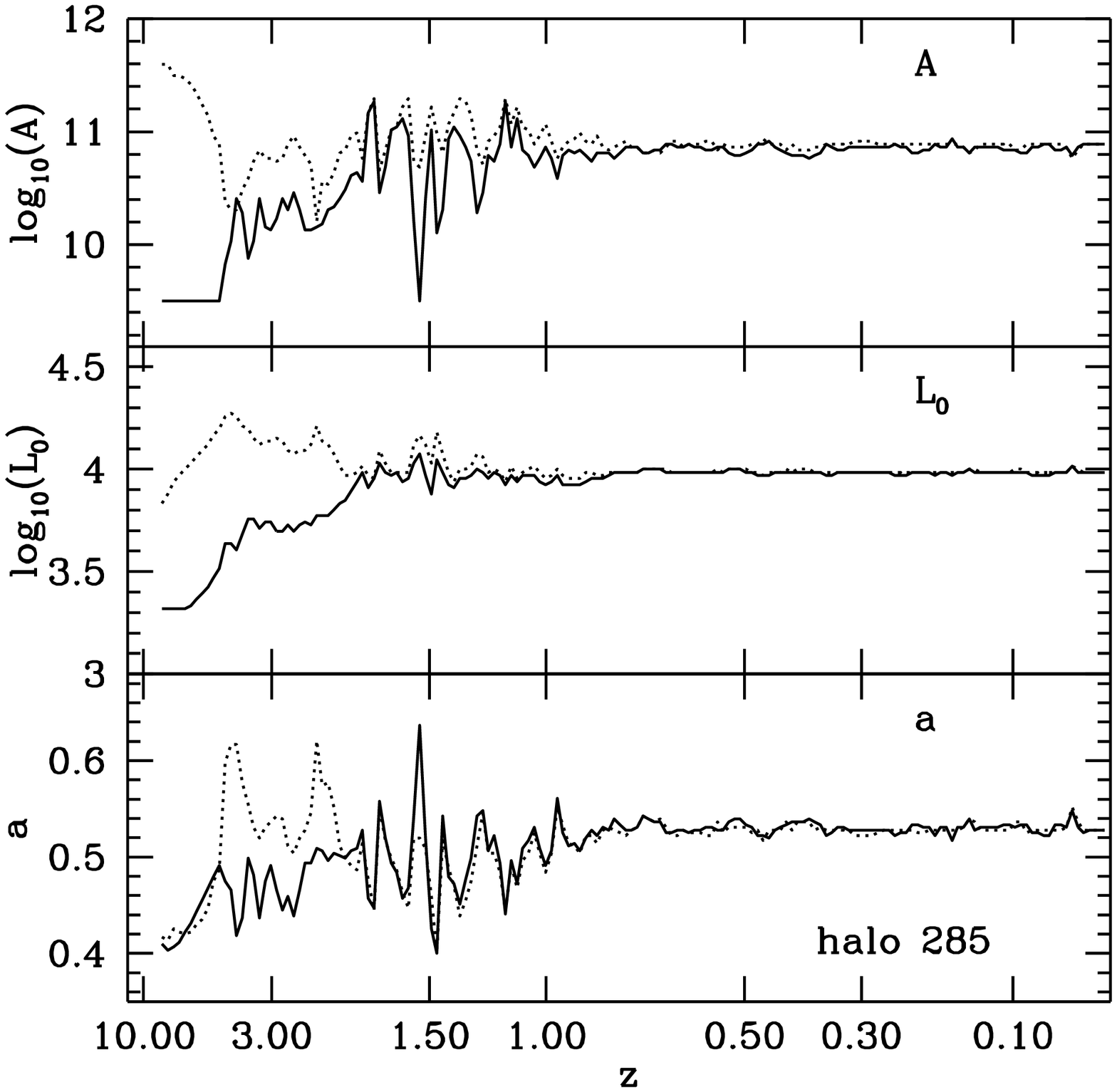}}
\caption{ Evolution of the parameters of $N = A\text{ }(\log_{10}(L/(\text{kpc km/s})) - \log_{10}(L_0))^2 \exp\left(-(\log_{10}(L/(\text{kpc km/s})) - \log_{10}(L_0))^2/2 a^2\right)$, the Boltzmann fitting function in Fig. \ref{fig:fits}, shown here for each of the four halos, where solid lines correspond to the particles selected to lie within $0.1\text{ }r_{vir}(z=0)$ as in the middle panel of Fig. \ref{fig:fits}, while the dotted lines correspond to the innermost $10^{11} M_{\odot}$ as shown in the bottom panel of Fig. \ref{fig:fits}. The $x$-axis shows the redshift. The trends in these plots are qualitatively related to the merger history of the halos.}
\label{fig:fitparams}
\vspace{200pt}
\end{figure*}

As a test, we show the fit parameters to the Boltzmann distribution as a function of redshift.  In Fig. \ref{fig:fitparams} we show the evolution of the parameters of the Boltzmann fitting function $N =\text{ }A (\log_{10}(L/(\text{kpc km/s})) - \log_{10}(L_0))^2 \exp\left(-(\log_{10}(L/(\text{kpc km/s})) - \log_{10}(L_0))^2/2 a^2\right)$, as shown in Fig. \ref{fig:fits}, for each of the four halos we consider. Solid lines correspond to the particles selected to lie within $0.1\text{ }r_{\rm vir}(z=0)$ in an Eulerian fashion, as in the middle panel of Fig. \ref{fig:fits}, while the dotted lines correspond to the innermost $10^{11} M_{\odot}$ as shown in the bottom panel of Fig. \ref{fig:fits}. The x-axis shows the redshift. Note that the discrete nature of the curves is a byproduct of the discrete sampling of fit parameters used to determine the minimum $\chi^2$ fit. The trends in these plots are qualitatively related to the merger history of the halos. We know that halo 277 (lower left) is quite quiescent since around $z=2$, which is reflected in the mostly smooth evolution of its fit parameters. Halo 258 (top right), whose last major merger occurred around $z=1$, shows roughly smooth evolution after this point. Halos 239 and 285 (top left and bottom right), which have more active merger histories up to the present time, have more variation in their parameter evolution.  As yet, though, we do not have a quantitative description of the fit parameters as a function of halo-evolution properties, nor do we know if a quantitative fit is possible.  Investigations into such a description would likely require a far larger statistical study than we have presented here.

\section{Causes of Angular Momentum Evolution}  \label{sec:why}

We have seen that there is significant evolution over the history of these halos of the average angular momentum (Sec. \ref{sec:Levol}) and of the angular momentum distribution (Sec. \ref{sec:Distrib}). We now investigate the origin of the torques that are causing this non-conservation of angular momentum. There are several possibilities for the origin of these torques, including external structure, non-sphericity of the halo itself, or infalling substructures that we would expect to play a role since we saw in the previous section that merger events are correlated with changes in the distribution of angular momenta.

\begin{figure*}
\centering
\subfloat[][]{\includegraphics[width=0.49\textwidth]{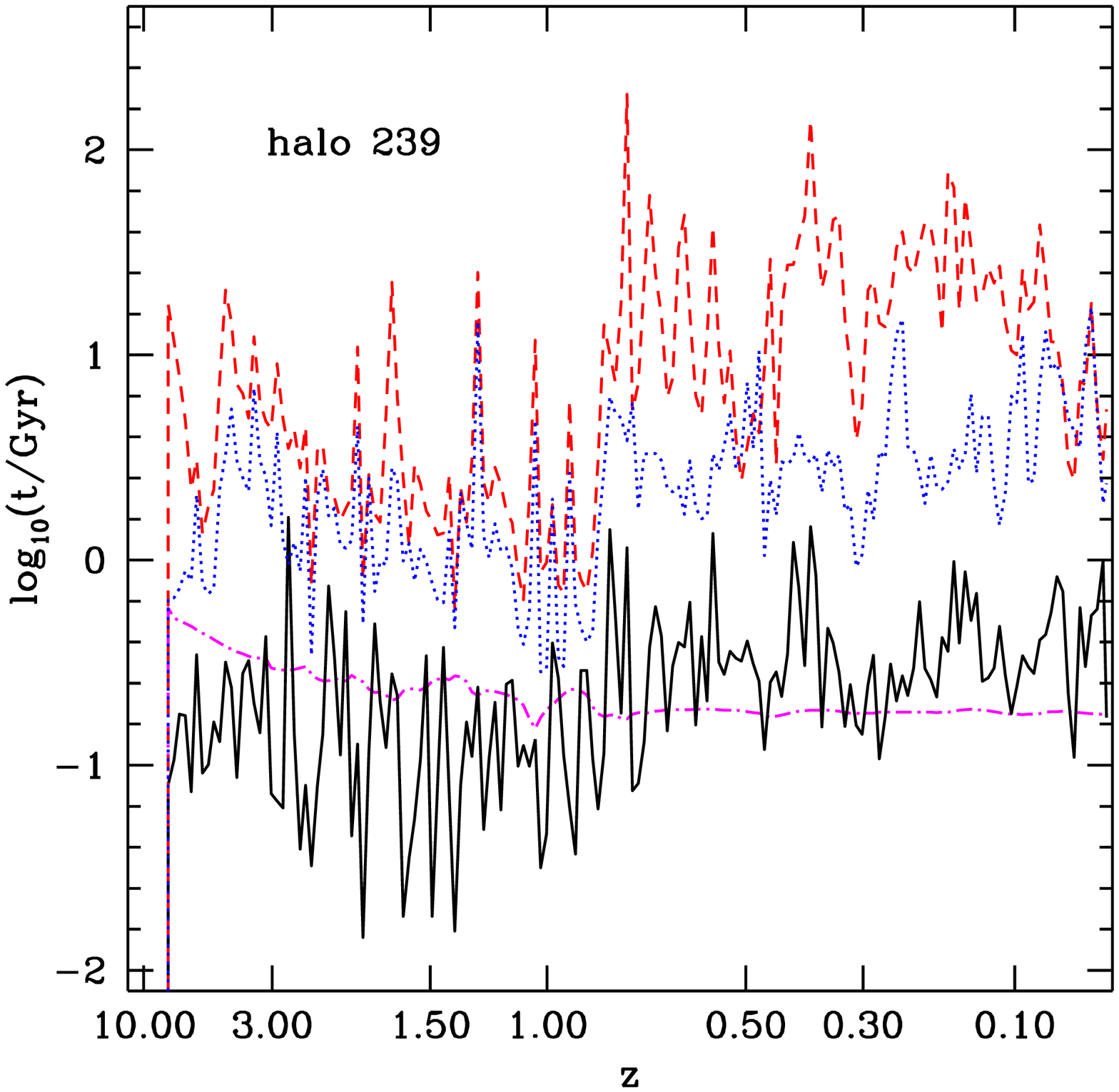}}
\subfloat[][]{\includegraphics[width=0.49\textwidth]{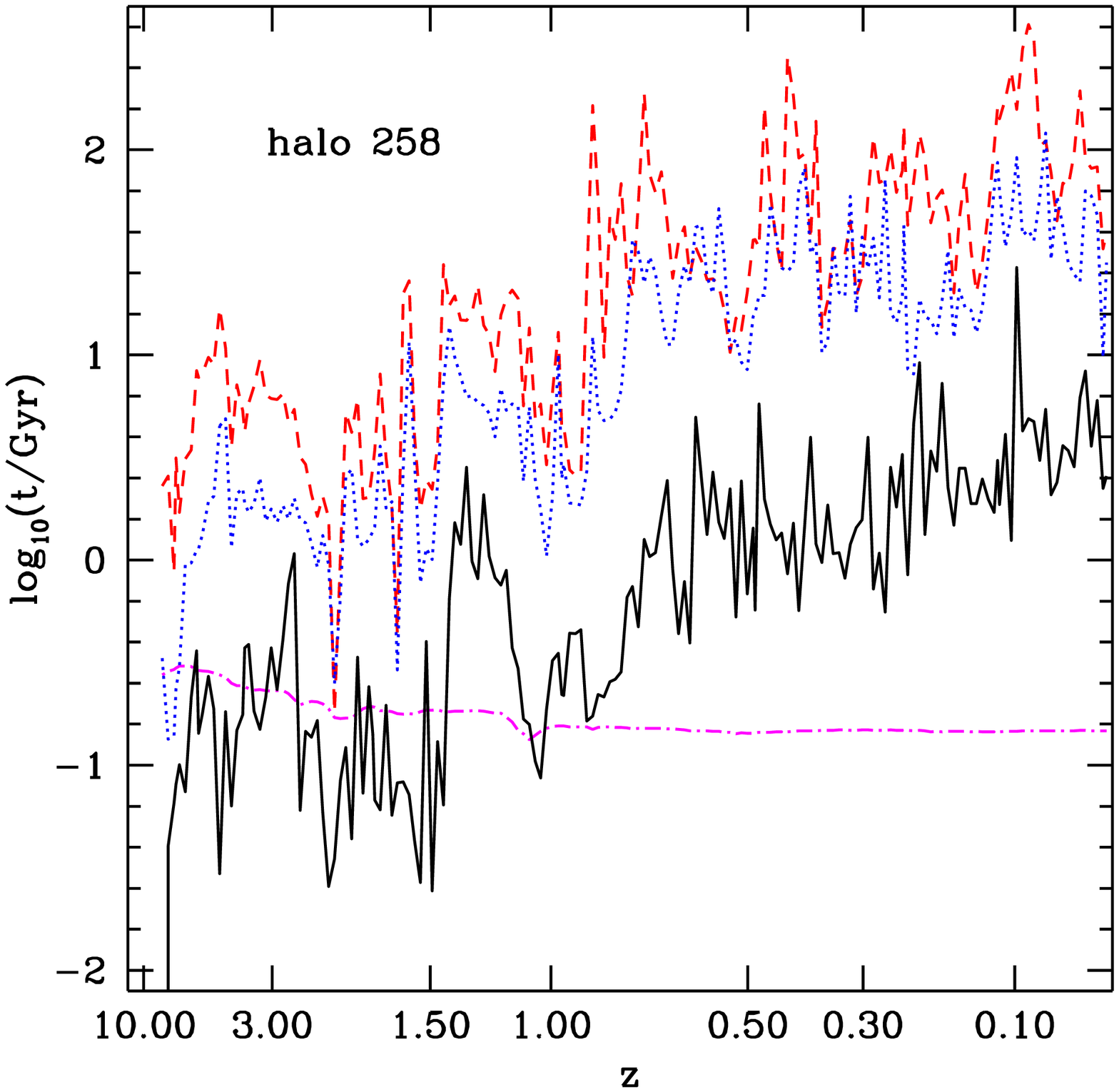}}\\
\subfloat[][]{\includegraphics[width=0.49\textwidth]{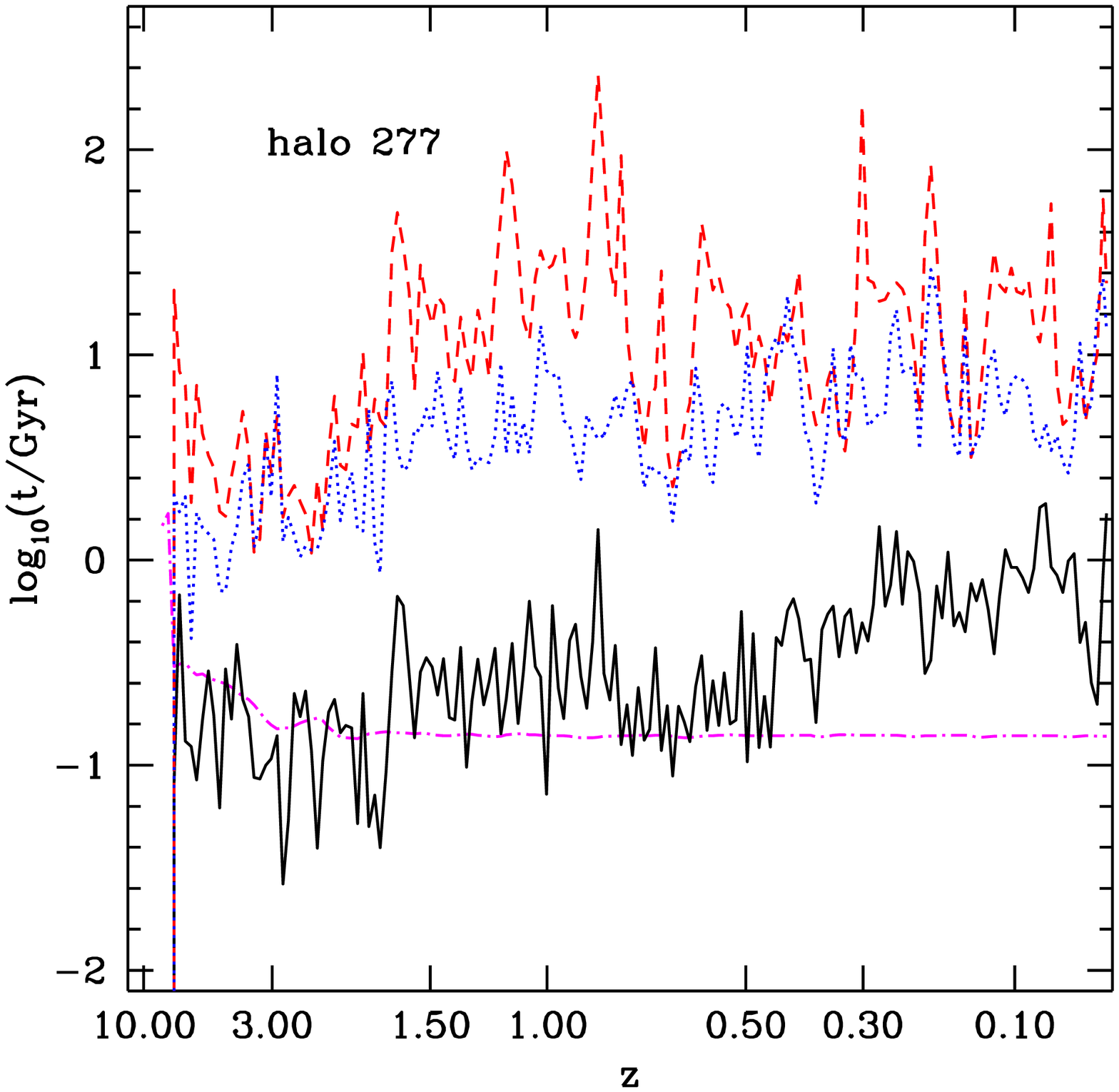}}
\subfloat[][]{\includegraphics[width=0.49\textwidth]{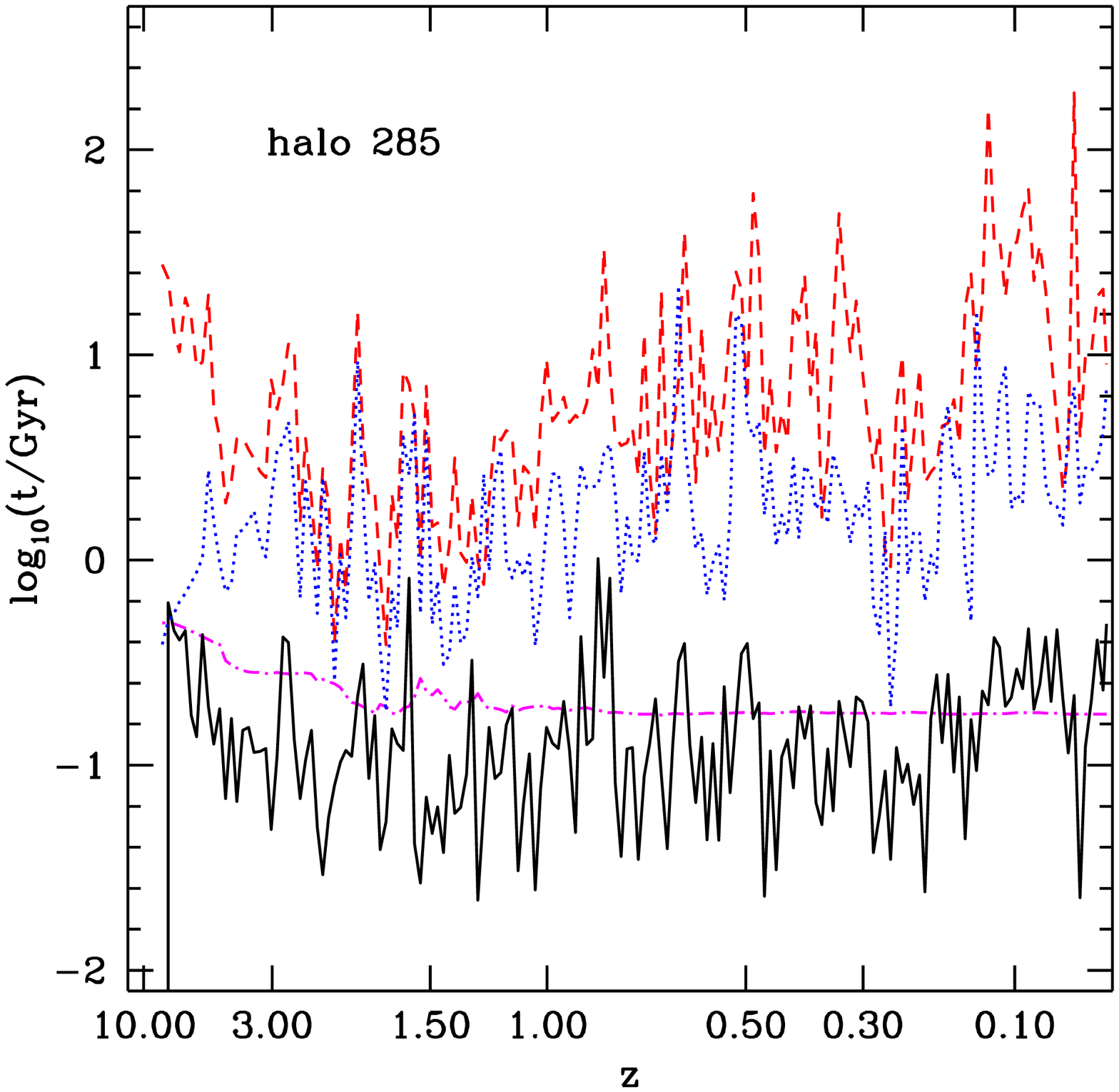}}
\caption{Time scales associated with torques on the inner halo due to outer halo particles (blue dotted lines), clumpy substructure particles (red dashed lines) and due to all torques (black solid lines), for halos (a) 239, (b) 258, (c) 277 and (d) 285. Also plotted is the inner halo dynamical time (magenta dot-dashed lines).}
\label{fig:timescales}
\vspace{200pt}
\end{figure*}

To investigate the relative importance of these mechanisms to the inner halo, we show in Fig. \ref{fig:timescales} the time scale associated with torques on the inner halo due to particles that are members of the host halo at $z=0$ (blue dotted lines) and due to particles which are members of clumped substructures which will be within the virial radius at $z=0$ (red dashed lines). We define the inner halo as those particles within the Eulerian-selected region $r(z) < 0.1\text{ }r_{\rm vir}(z=0)$. We compare this with the time scale for the angular momentum to change due to all torques, defined as the ratio of the total angular momentum of the inner region to its time derivative $L_{\rm in}/[{\rm d}L_{\rm in}/{\rm d}t]$ (black solid lines), and the inner halo dynamical time (magenta dot-dashed lines). Note that a longer time scale here implies a weaker torque, and conversely a small time scale implies a strong torque that can change the angular momentum relatively quickly.

From these figures we can see that, in general, the time scale for the angular momentum to change by order unity due to all external and internal torques is at least an order of magnitude less than that for the angular momentum to change due to torques only by other particles in the halo or by clumped substructures, implying that the torques due to non-sphericity of the host halo and clumpy substructure are far weaker than torques due to external structure in the angular momentum evolution of the inner halo. Thus, external torques are the dominant source of angular momentum evolution of the inner halo as a whole. We also observe that all of these time scales become large relative to the inner halo dynamical time during quiescent merger phases, for example after redshift $z\sim 1$ for halo 258, while the angular momentum change time scale becomes comparable to the dynamical time during mergers both major and minor. This is reasonable, as it implies that the angular momentum changes slowly during periods of quiescent evolution. Note that the angular momentum change time scale of halo 277 is closer to the dynamical time scale during its quiescent evolution after $z=3$, while the angular momentum change time scale of halo 258 becomes significantly larger than its dynamical time scale when it is quiescent after $z\sim1$. This indicates that the speed of the evolution of angular momentum during quiescent phases does vary between halos, probably due to the environment of the halo which is producing external torques.

\begin{figure*}
\centering
\subfloat[][]{\includegraphics[width=0.33\textwidth]{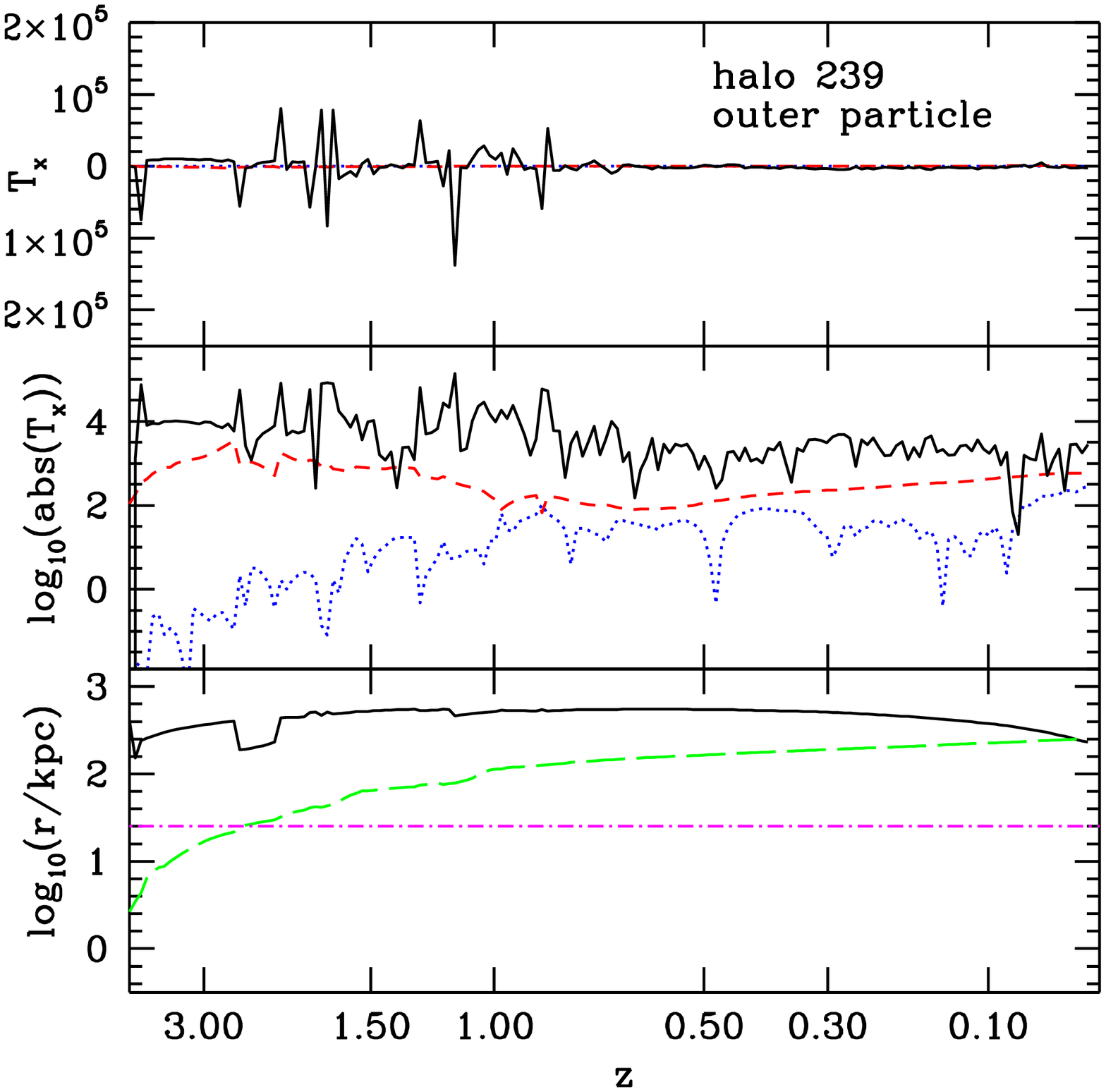}}
\subfloat[][]{\includegraphics[width=0.33\textwidth]{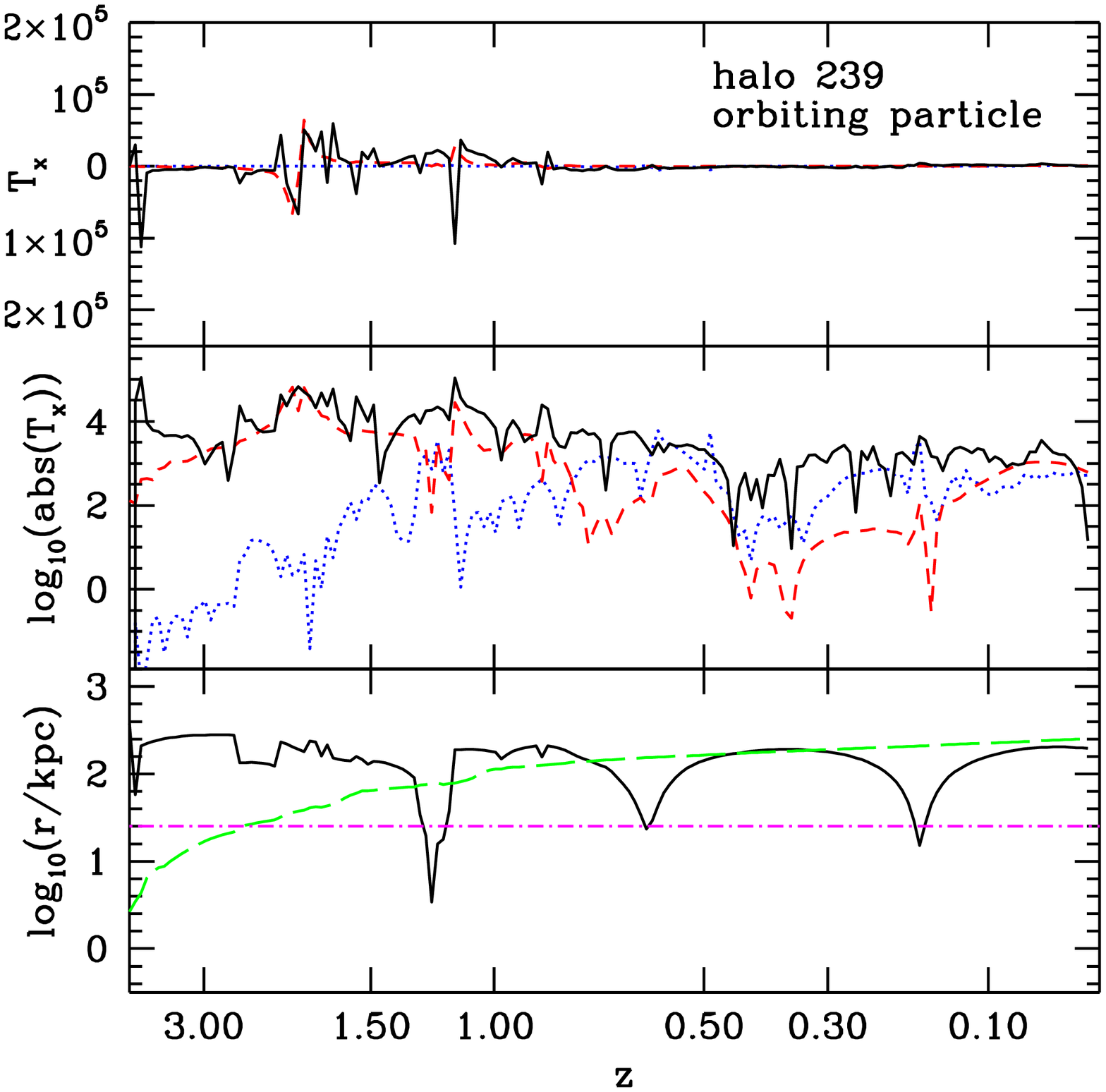}}
\subfloat[][]{\includegraphics[width=0.33\textwidth]{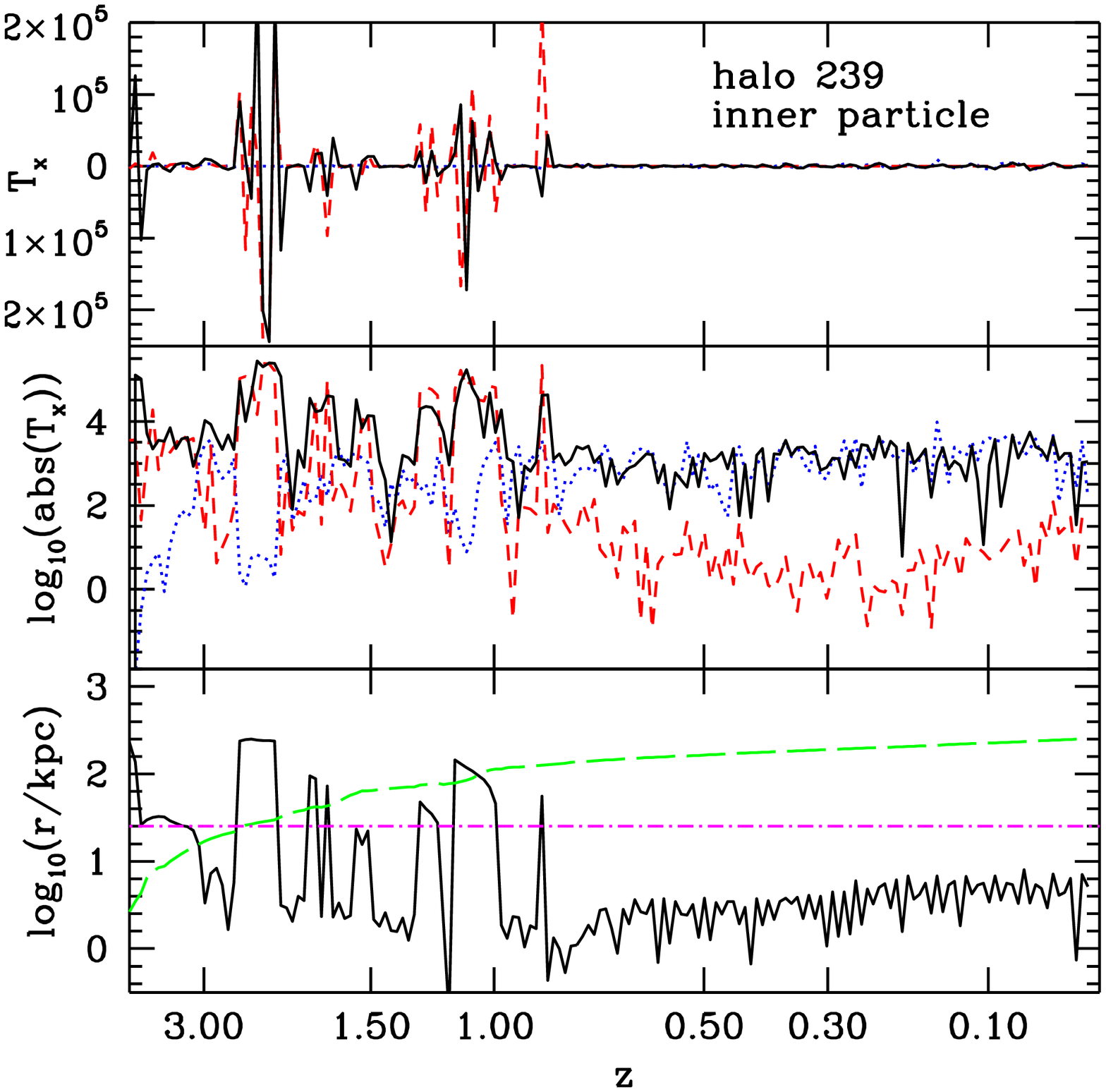}}\\
\subfloat[][]{\includegraphics[width=0.33\textwidth]{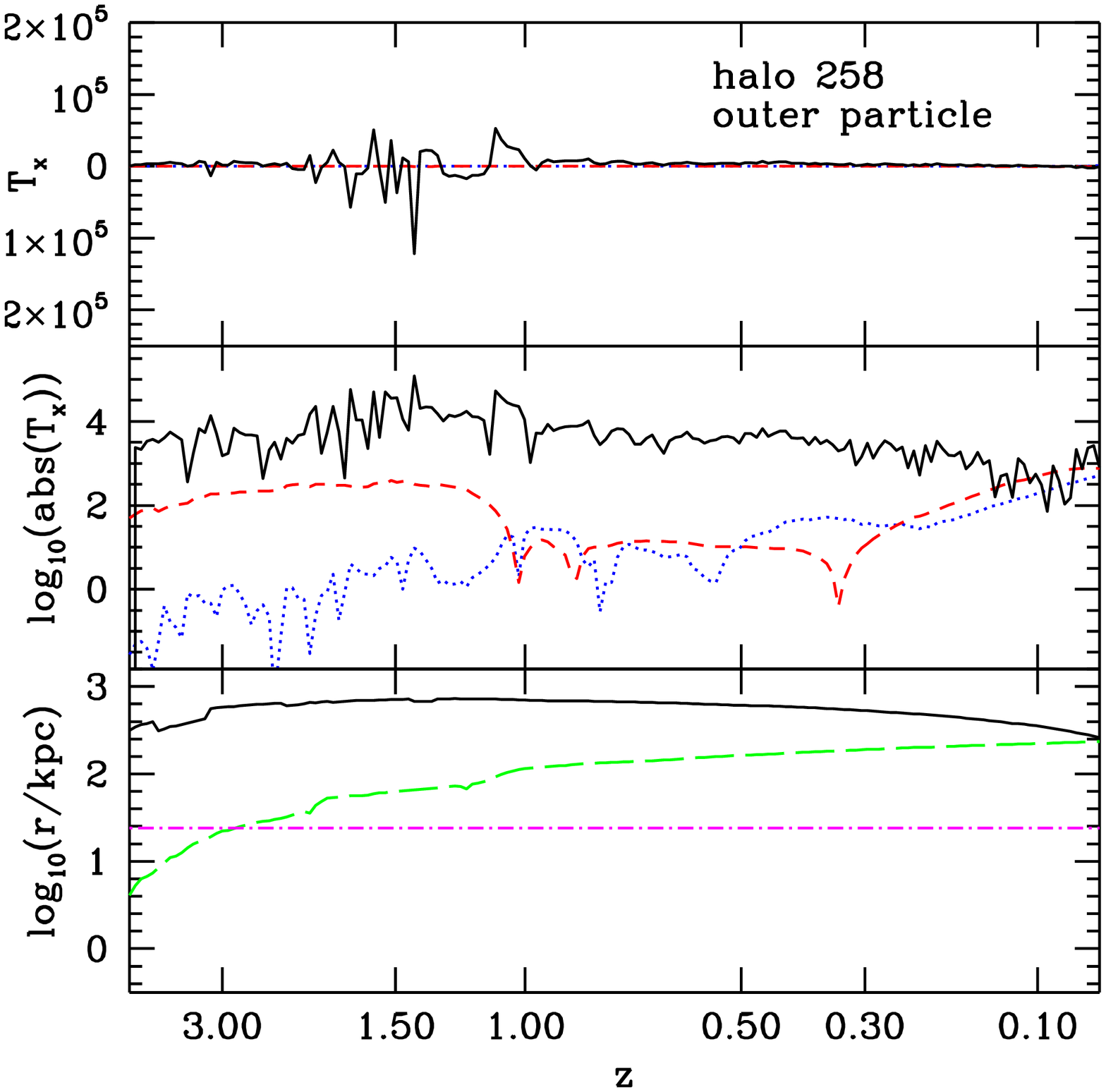}}
\subfloat[][]{\includegraphics[width=0.33\textwidth]{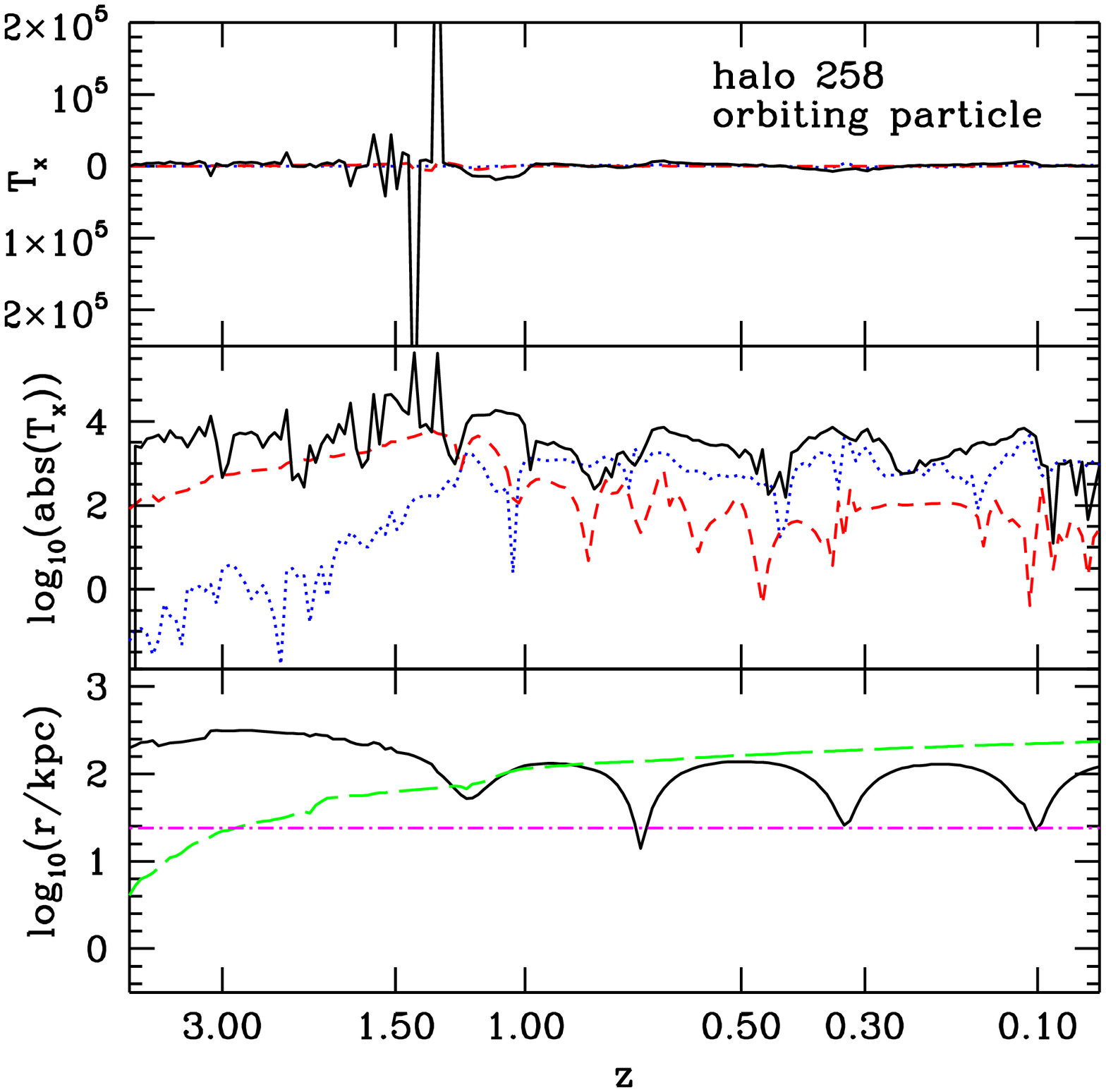}}
\subfloat[][]{\includegraphics[width=0.33\textwidth]{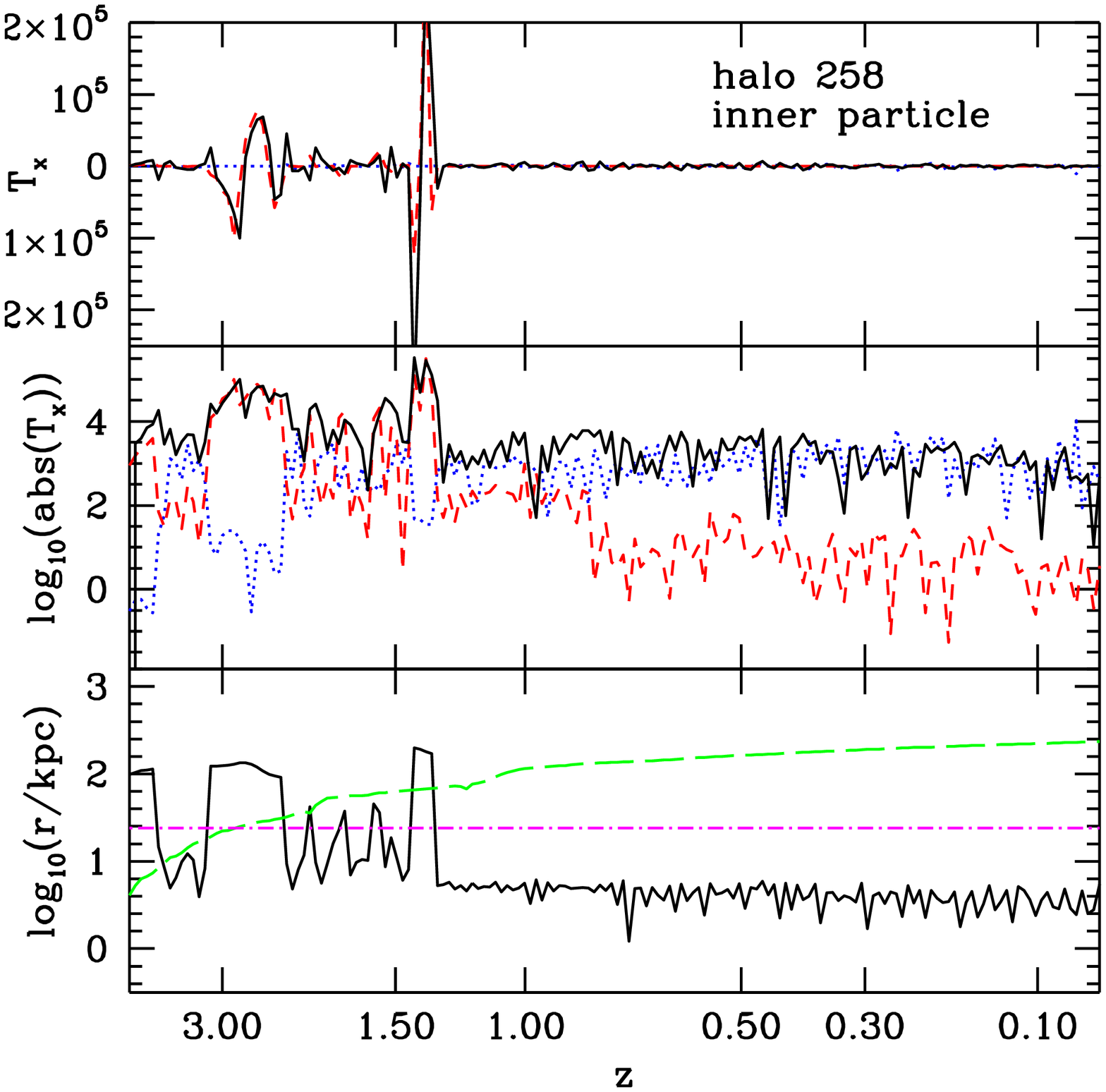}}
\caption{Total torque on selected halo particles (solid black lines) compared with the torque due to halo particles (blue dotted lines) and particles which are members of clumped substructures that will be within the virial radius of the halo at $z=0$ (red dashed lines). The torque is shown on a linear scale in the top panel, and its absolute value is shown on a log scale in the middle panel of each figure. The units of the torque in the top two panels are km kpc s$^{-1}$ Gyr$^{-1}$. In the bottom panel, we show the radius of this particle (black solid lines) and compare it with the virial radius of the halo (green long-dashed lines), and the line $0.1\text{0}r_{\rm vir}(z=0)$ (magenta dot-dashed lines). The top three plots show three particles in halo 239, while the bottom three are from halo 258. For each halo we have chosen one ``outer", one ``orbiting" and one ``inner" particle to show here.}
\label{fig:clumps}
\vspace{200pt}
\end{figure*}

We address the role played by these torques on individual particles in Fig. \ref{fig:clumps}, in which we show the total torque ${\rm d}L/{\rm d}t$ on selected particles (solid black lines), and compare this with the torque on these particles due to particles that are currently bound to the main halo and no other substructure (blue dotted lines) and particles which are members of clumped substructures at the given redshift that will be within the virial radius of the halo at $z=0$ (red dashed lines). These torques are shown on a linear scale in the top panel, and their absolute value on a log scale is shown in the middle panel of each figure. The difference between the sum of the torques due to halo particles and clumped particles and the total torque is the torque due to external structure, from particles that are not members of the halo at $z=0$. The units of the torque in the top two panels are km kpc s$^{-1}$ Gyr$^{-1}$. In the bottom panel, we show the radius of each individual particle (black solid lines) and compare it with the virial radius of the halo (green long-dashed lines), and the line $0.1\text{0}r_{\rm vir}(z=0)$ (magenta dot-dashed lines) below which particles are classified as inner halo particles. The top three plots show three particles in halo 239, while the bottom three are from halo 258. For each halo we have randomly chosen one particle that spends most of its time outside the virial radius (``outer particle"), one that orbits the halo, entering the virial radius on its closest approach (``orbiting particle"), and a particle that spends much of its time in the inner regions of the halo (``inner particle").

In general, we see that the torque on the individual particles that we have chosen is dominated by smaller subhalos before the particle is accreted, but that this torque falls below that due to halo particles after the particle is accreted onto the halo. However, in Fig. \ref{fig:timescales} we saw that the angular momentum of the inner halo as a whole is dominated by external torques, with mergers also tending to lower the inner halo angular momentum. When taking the individual particle behaviour shown in Fig. \ref{fig:clumps} together with the behaviour of the entire inner halo in Fig. \ref{fig:timescales}, we see that the importance of torques due to halo and clumped particles in the evolution of some of the individual particles goes away when considering the inner halo as a whole. Also, external structure, which plays a relatively minor role in the evolution of some of the individual particle angular momenta, is a dominant effect on the evolution of the total angular momentum of the inner halo. We may interpret this to say that external structure tends to torque each particle in the entire halo coherently, while torques due to other halo particles and clumpy substructures can be larger on individual particles but are incoherent when considering the entire inner halo. If this is the case, then the torques due to halo particles will tend to cancel out when averaged over many individual particles in the inner halo, explaining the relative unimportance of these torques for the entire inner halo, and why external structure comes to dominate the evolution of the inner halo.

\section{Discussion}  \label{sec:disc}

We have analyzed whether the assumptions of the AC model are valid in the case of simulated dark matter halos. Specifically, we have looked into whether such halos can be characterized as adiabatically evolving, and whether their angular momentum or angular momentum distributions are conserved as we would expect from the assumption of spherical symmetry in the model of AC most commonly applied to dark matter halos. We have found that the assumption of adiabatic evolution of halos is not completely correct, that the angular momentum of regions of the halo is not an adiabatic invariant as is assumed in the AC model, and that its distribution also varies. We have investigated the sources of these torques, and found that external torques are the main source of torque on the halo as a whole.

In greater detail, our major results are as follows:

\begin{enumerate}
\item Halo particles follow the trend of centre-of-mass angular momentum expected from tidal torque theory.

\item In general, the gravitational potential of our halos changes on time scales larger than the dynamical time in the halos, which is a prerequisite for adiabatic invariance. For the halo as a whole, the potential change time scale is generally around an order of magnitude larger than the dynamical time scale, while for the inner halo the potential change time scale is only a  factor of a few larger than the dynamical time scale.

\item Angular momentum is lost from all particles as halos virialize, more from those particles that end up in the centre of halos. We find that both the vector- and magnitude-averaged angular momentum in fixed radial bins about the halo centre decreases with time, by a few tens of percent to factors of a few. The vector-averaged quantities usually decrease more than the magnitude-averaged ones, implying that the directions of angular momenta in radial bins become progressively more misaligned over time. 

\item The distribution of angular momentum magnitudes is well fit by a simple Boltzmann fitting function. Trends in the evolution of these distributions qualitatively reflect the merger history of the halo.

\item External torques dominate the angular momentum evolution of the inner halo, while substructure and halo non-sphericity torques can be important for the angular momentum evolution of individual particles. The dominant role of external torques in changing the angular momentum of halo particles agrees with the results of \citet{Valluri10}, who found the evolution of halos during baryonic condensation to be mostly reversible when external torques were not included.
\end{enumerate}

We find that halo particles are losing angular momentum even in these dark-matter-only simulations, and that the net angular momentum loss of the inner regions of the halo is due mainly to external tidal torques. The amount of this decrease depends on what region we choose and how we add the angular momenta, but the ratio of angular momenta at $z=1$ and $z=0$ can vary from a few tenths to a few.  The distribution of angular momenta in both the halos as a whole and in the innermost parts in which a galaxy would live are not time-invariant, which means that the spherically-symmetric form of adiabatic contraction that is typically applied to dark-matter halo profiles is not strictly valid.  

While we find indications that the evolution in the angular-momentum distribution is correlated with the halo accretion histories, we have not found a quantitative description of these changes.  Larger statistical studies are required to determine if a quantitative relation between the angular momentum distribution and halo accretion history can be established.

Many previous studies of AC have found it to overpredict the effect of baryon condensation on dark matter density in the centres of halos under the assumption that angular momentum is conserved. Our finding that angular momentum is lost from all particles in the halo over time likely exacerbates this problem, as it implies that even more mass would be concentrated in the centre due to dark matter only interactions. The interaction of the angular momentum loss of dark matter particles observed here with baryonic physics is an interesting direction for future research.

That angular momentum does not behave as a perfect adiabatic invariant is not surprising since, for example, halos are non-spherical and evolve with time. Our results serve to highlight the magnitude of this issue and serve as a caution to applications of the adiabatic contraction approximation---there is a limit to the precision which we can reasonably expect it to provide. The fact that recent simulations show that no single model of adiabatic contraction works well in all cases (e.g. \citealt{Gustafsson06}) suggests that this limit may have been reached. Overall, our results imply that a fundamental limit to the applicability of current adiabatic contraction models which should be kept in mind when applying these approximations to the effects of galaxy formation on dark matter halos.

\section*{Acknowledgments}
LGB acknowledges the support of the NSF Graduate Fellowship Program. AHGP and AJB are supported by the Gordon and Betty Moore Foundation. AMB is supported by the Sherman Fairchild Foundation.  All simulations were run with resources at the NASA Advanced Supercomputing Division.

%
%

\bibliographystyle{mn2e} 
\bibliography{ref}
%
%


\label{lastpage}
\end{document}